\documentclass[lettersize,journal]{IEEEtran}
\usepackage{amsmath,amsfonts}
\usepackage{array}
\usepackage[caption=false,font=normalsize,labelfont=sf,textfont=sf]{subfig}
\usepackage{textcomp}
\usepackage{stfloats}
\usepackage{url}
\usepackage{verbatim}
\usepackage{graphicx}
\usepackage{cite}
\usepackage{color}
\usepackage{amssymb}
\usepackage{amsthm}
\usepackage{bbding}
\usepackage{mathrsfs}
\usepackage[ruled,linesnumbered]{algorithm2e}

\newtheorem{definition}{Definition}
\usepackage{multirow}
\usepackage{epstopdf}
\usepackage{epsfig}
\usepackage{arydshln} 
\begin{document}

\title{TSGN: Transaction Subgraph Networks Assisting Phishing Detection in Ethereum}
\author{Jinhuan Wang,
        Pengtao Chen,
        Xinyao Xu,
        Jiajing Wu,~\IEEEmembership{Senior Member,~IEEE},
        Meng Shen,~\IEEEmembership{Member,~IEEE}, \\
        Qi Xuan,~\IEEEmembership{Senior Member,~IEEE},
        Xiaoniu Yang

\IEEEcompsocitemizethanks{
\IEEEcompsocthanksitem J. Wang, P. Chen, and X. Xu are with the Institute of Cyberspace Security, College of Information Engineering, Zhejiang University of Technology, Hangzhou 310023, China. E-mail: jhwang@zjut.edu.cn; Pengt.Chen@gmail.com; 654202923@qq.com.
\IEEEcompsocthanksitem J. Wu is with the School of Computer Science and Engineering, Sun Yat-sen University, Guangzhou 510006, China. E-mail: wujiajing@mail.sysu.edu.cn.
\IEEEcompsocthanksitem M. Shen is with the School of Cyberspace Science and Technology, Beijing Institute of Technology, Beijing 100081, China, and also with Peng Cheng Laboratory (PCL), Shenzhen 518066, China. E-mail: shenmeng@bit.edu.cn.
\IEEEcompsocthanksitem Q. Xuan is with the Institute of Cyberspace Security, College of Information Engineering, Zhejiang University of Technology, Hangzhou 310023, China, with the PCL Research Center of Networks and Communications, Peng Cheng Laboratory, Shenzhen 518000, China, and also with the Utron Technology Company Ltd., Hangzhou 310056, China. E-mail: xuanqi@zjut.edu.cn.
\IEEEcompsocthanksitem X. Yang is with the Institute of Cyberspace Security, Zhejiang University of Technology, Hangzhou 310023, China, and also with the Science and Technology on Communication Information Security Control Laboratory, Jiaxing 314033, China. E-mail: yxn2117@126.com.
\IEEEcompsocthanksitem Corresponding authors: Qi Xuan.}
}





\maketitle

\begin{abstract}

Due to the decentralized and public nature of the Blockchain ecosystem, the malicious activities on the Ethereum platform impose immeasurable losses for the users.
Existing phishing scam detection methods mostly rely only on the analysis of original transaction networks, which is difficult to dig deeply into the transaction patterns hidden in the network structure of transaction interaction.
In this paper, we propose a \underline{T}ransaction \underline{S}ub\underline{G}raph \underline{N}etwork (TSGN) based phishing accounts identification framework for Ethereum. We first extract transaction subgraphs for target accounts and then expand these subgraphs into corresponding TSGNs based on the different mapping mechanisms. In order to make our model incorporate more important information about real transactions, we encode the transaction attributes into the modeling process of TSGNs, yielding two variants of TSGN, i.e., Directed-TSGN and Temporal-TSGN, which can be applied to the different attributed networks. Especially, by introducing TSGN into multi-edge transaction networks, the Multiple-TSGN model proposed is able to preserve the temporal transaction flow information and capture the significant topological pattern of phishing scams, while reducing the time complexity of modeling large-scale networks. Extensive experimental results show that TSGN models can provide more potential information to improve the performance of phishing detection by incorporating graph representation learning.

\end{abstract}




\section{Introduction}\label{sec:introduction}
\IEEEPARstart{T}{hrough} consensus mechanisms and voluntary respect for the social contract, the internet can be utilized to make a decentralised value-transfer system that can be shared around the world and virtually free to use~\cite{nakamoto2008bitcoin}. Ethereum~\cite{wood2014ethereum}, a very professional system of a cryptographically secure, offers us information that has never been seen before in the financial world. In contrast to fiat currencies of financial world, the transactions through virtual currencies like Ether are completely transparent. Accounts can freely and conveniently conduct transactions with currency and information and do not have to rely on traditional third parties, and these transactions of cryptocurrencies are permanently recorded on Blockchain and are available at any time.


There is no doubt that Ethereum platform facilitates transactions between consenting individuals who are trust with each other. But, the cryptocurrency market provided by Ethereum is inevitably flooded with a variety of cybercrimes due to anonymity and unsupervised organization, including smart Ponzi schemes, phishing, money laundering, fraud, and criminal-related activities. The overall losses caused by DeFi exploits on Ethereum, the backbone of many DeFi applications, have totaled \$12 billion so far in 2021 according to Elliptic~\footnote{\url{https://blockchaingroup.io/}}, a firm which tracks movements of funds on the digital ledgers that underpin cryptocurrencies. Moreover, it is reported that phishing scams can break out periodically and are the most deceptive form of fraud~\cite{chen2020phishing}. In general, phishing is a kind of cybercrime aiming to exploit the weaknesses of users and obtain personal and confidential information~\cite{khonji2013phishing}. Some scams lure investors to visit/click the fake projects/websites by offering extra Ether coins as the enticement, and to buy digital assets from scammers for the purpose of phishing scams. Although the hash mechanism set up inside the blockchain can prevent transactions from being tampered with, so far, there are no available internal tools that can detect illegal accounts and suspicious transactions on the network. Thus it can be seen that cybercrimes, especially phishing scams, have become a critical issue on Ethereum and should be worthy of long-term attention and research to maintain a more secure blockchain ecosystem.

Existing methods of phishing detection on Ethereum mainly fall into two categories, propose elaborate handcrafted features~\cite{huang2020understanding,chen2020phish,poursafaei2020detecting} and design automatical feature learning models~\cite{hu2021scsguard,wu2020phishers,alarab2020competence}.
These studies attempt to learn more useful information from the raw transaction data to reveal the transaction patterns, which are effective for phishing detection but have several important omissions. First, the transaction networks constructed by sampling the original Ethereum transaction data may result in missing structural information which can affect the performance of learning models. 
This forces researchers to design more precise and data-targeted representation methods to improve the effect, but at the expense of the generalization performance of models on datasets of different attributes and scales. 
Second, the above models are well-designed only for learning the original networks, ignoring the potential transaction flow patterns provided by the interaction of transactions.

\begin{figure}[!t]
    \centering
    \includegraphics[width=1\linewidth]{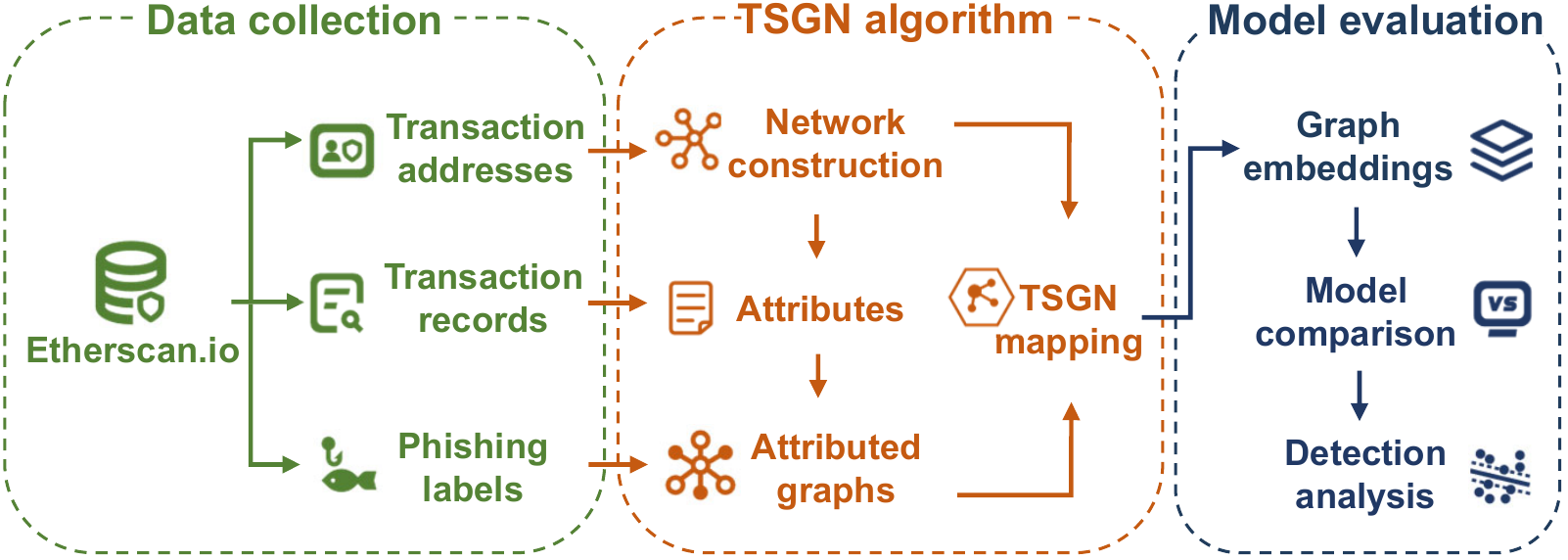}
    \caption{The framework of our TSGN algorithm for phishing detection on Ethereum.}
    \label{fig:framework}
\end{figure}

In this paper, inspired by SGN model~\cite{xuan2019subgraph}, we develop a network model called TSGN to complement the features of original networks and further assist in identifying phishing accounts. According to the essential attributes contained in the transactions, we design the specific structural mapping strategies in order to encode the weight, direction and timestamp information into transaction subgraph networks. And then, the transaction network is mapped into the transaction subgraph structure space according to the different strategies to obtain the corresponding TSGN and its variants for the subsequent feature extraction and detection task. The algorithm framework is shown in Fig.~\ref{fig:framework}. Specifically, our contributions can be concluded as follows:


\begin{itemize}
\item We propose a novel transaction network model, transaction subgraph networks (TSGNs). According to original transaction networks (TN) of various attributes and scales, our TSGNs can introduce different network mapping strategies to fully capture the potential structural topological information which can not be obtained easily from raw transaction networks, benefiting the subsequent phishing detection algorithms.
\item We built the first variant of TSGN model, i.e., Directed-TSGN, to integrate direction attribute for effectively characterize transaction flow information from the original transaction data. And, we further introduce timestamp information into Directed-TSGN, producing the Temporal-TSGN which can exactly learn the time-aware transaction behaviors and is applied on multi-edge transaction networks.
\item Experimental results demonstrate the effectiveness of the TSGN models by integrating various phishing detection models such as Handcrafted features, Graph2Vec, DeepKernel, Diffpool, and U2GNN. Especially, the classification result of Temporal-TSGN increases to 98.43\% (97.00\% for Directed-TSGN) when only Diffpool is considered, greatly improving the phishing detection performance. More remarkably, compared with TSGN, the variants of TSGN are of the controllable scale and have lower time cost.
\end{itemize}

The rest of the paper is structured as follows. In section \ref{sec:related}, we make a brief description of the phishing identification and graph representation methods. In Section~\ref{sec:method}, we mainly introduce the definitions and construction methods of transaction subgraph networks. In Section~\ref{sec:experiment}, we give several feature extraction methods, which together with TN, TSGN, Directed-TSGN, and Temporal-TSGN are applied to six Ethereum transaction network datasets. Finally, we conclude our paper in Section~\ref{sec:conclusion}. For further study, our source codes are available online~\footnote{\url{https://github.com/GalateaWang/TSGN-master}}.


\section{Background and Related Work}\label{sec:related}

In this section, we provide some necessary background information on phishing detection methods and graph representation algorithms for blockchain data mining.

\subsection{Phishing Identification} \label{subsec:phishident}

To create a good investment environment in the Ethereum ecosystem, many researchers have paid lots of attention to study the effective detection methods for phishing scams.
Wu et al.~\cite{wu2020phishers} proposed an approach to detect phishing scams on Ethereum by mining its transaction records. By considering the transaction amount and timestamp, it introduced a novel network embedding algorithm called trans2vec to extract the features of the addresses for subsequent phishing identification.
Chen et al.~\cite{chen2020phish} proposed a detecting method based on Graph Convolutional Network (GCN) and autoencoder to precisely distinguish phishing accounts.
Li et al.~\cite{li2021self} introduced a self-supervised incremental deep graph learning model based on Graph Neural Network (GNN), for phishing scam detection problem on Ethereum.
One can see that these methods mentioned above mainly built phishing account detection as a node classification task, which cannot capture more potential global structural features for phishing accounts.
Yuan et al.~\cite{yuan2020phishing} built phishing identification problem as the graph classification task, which used line graph to enhance the Graph2Vec method and achieved good performance. However, Yuan et al. only consider the structural features obtained from line graphs, ignoring the transaction temporal information which plays a significant role in phishing scam detection.
As we know, the phishing funds mostly flow from multiple accounts to a specific account. 
Our method takes the direction and temporal information into consideration and builds the Directed-TSGN and Temporal-TSGN models, revealing the topological pattern of phishing scams.

\subsection{Graph Representation on Blockchain} \label{subsec:graph}


Graphs are a general language for describing and analyzing entities with relations or interactions. Due to its unique structure characteristics, the Blockchain ecosystem is naturally modeled as transaction networks to carry out related research. Simultaneously, many graph representation methods are applied to capture the dependency relationships between objects in the Blockchain network structures.
There are two main categories of existing representation methods, the former mostly employs machine learning models with dedicated feature engineering~\cite{huang2020understanding,chen2020phish,poursafaei2020detecting}, and the latter adopts some embedding techniques and deep learning models such as walk-based~\cite{wu2020phishers} and the GNN methods. For example, Hu et al.~\cite{hu2021scsguard} presented a novel deep learning scam detection framework assembling with the Gate Recurrent Unit (GRU) network and attention mechanism, which produced good results on both Ponzi and Honeypot scams.
Alarab et al.~\cite{alarab2020competence} adopted GCNs intertwined with linear layers to predict illicit transactions in the Bitcoin transaction graph and this method outperforms graph convolutional methods used in the original paper of the same data.
Liu et al.~\cite{liu2020graph} introduced an identity inference approach based on big graph analytics and learning, aiming to infer the identity of Blockchain addresses using the graph learning technique based on GCNs.
Zhang et al. \cite{zhuang2020smart} constructed a graph to represent both syntactic and semantic structures of an Ethereum smart contract function and introduced the GNN for smart contract vulnerability detection.

According to the above works, one can find that the graph representation methods, especially GNNs, can indeed be utilized to study the Blockchain data and outperform in many different applications.
However, the existing studies have two potential problems. Firstly, these researches are mostly based on a large-scale transaction network where the scarcity of labels and the huge volume of transactions make it difficult and intricate to take advantage of GNN methods. Secondly, such methods largely rely on the original transaction network and thus may overlook the powerful potential of transaction interactions which contain important hidden transaction patterns.
In this work, we collect the 1-hop neighborhood information to construct the subgraph for each target address and use the features of the entire subgraph as the representation of the target node for avoiding inadequate graph feature learning. We further obtain the TSGNs and their three variants by structural space mapping, capturing potential transaction patterns from various aspects. 

\section{Methodology}\label{sec:method}

We first give the description of problem formulation of phishing identification and then illustrate the construction details of the transaction subgraph network models by introducing several definitions.

\subsection{Problem Formulation} \label{subsec:problem}

Generally, a cryptocurrency transaction network is modeled as an address interaction graph, illustrating the transferring amount and the flow direction of fund between different accounts in a certain time bucket. Given a set of addresses on Ethereum, we can construct transaction network as a directed graph $G=(V,E)$, where each node $v\in V$ indicates an address that can be an \emph{Externally Owned Account} (EOA) or \emph{Contract Account} (CA), the edge set is represented as a quad, $E = \{(v_{src},v_{dst},w,\ell)|v_{src},v_{dst} \in V, w \in \mathbb{R}^+ \cup \{0\}, \ell \in L\}$, where ($v_{src},v_{dst}$) indicates the direction of the transaction (e.g., assets transfer or smart contract invocation) from $v_{src}$ to $v_{dst}$, weight $w$ represents the amount of transferred cryptocurrencies or is 0 if the kind of transaction is smart contract invocation, and $l$ is the timestamp of the transaction.


Here, we construct a set of transaction graphs for the center addresses $\mathcal{G} = \{(G_1,c_1),(G_2,c_2),\cdots,(G_n,c_n)\}$, where $c_i \in C^{|\mathcal{G}|\times |\phi|}$ is the label of address $i$ and its corresponding transaction subgraph $G_i$, where $\phi$ is the label set of all target addresses. In this work, our purpose is to learn a mapping function $\mathscr{C}:\mathcal{G} \rightarrow C$ which can predict the labels of graphs in $\mathcal{G}$. The label set $C$ includes phishing labels and non-phishing labels in the scenario of Ethereum phishing account identification. 


To keep mathematical formulas unified in a stricter and tidier manner, TABLE~\ref{notation} shows the notations used in the progress of TSGNs modeling.

\begin{table}[t]
\renewcommand{\arraystretch}{1.2}
\caption{Mathematical notations used in this paper.}
\label{notation}
\centering
  \begin{tabular}{lr}
    \hline\hline
    Notation & Description \\
    \hline
    $G$ & The original Ethereum transaction graph \\
    
    $T$ & The TSGN graph \\
    
    $T_D$ & The Directed-TSGN graph \\
    
    $T_T$ & The Temporal-TSGN graph \\
    
    $T_{\mathcal{A}}$ & The generic notation of TSGN graph with an attribute \\
    
    $v$, $u$ & The account nodes in graph $G$ \\

    $e$ & A transaction edge in graph $G$ \\

    $d$ & The directed transaction edge in graph $G$ \\

    $t$ & The temporal directed transaction edge in graph $G$ \\
    
    $\ell$ & The timestamp attached on the edge $t$\\
    
    $w$, $w'$, $w^{*}$, $w_{\times}$ & The edge weights set by the transaction amount \\
    
    $V$, $V'$, $V^{*}$, $V_{\times}$ & The node sets \\

    $E$, $E'$, $E^{*}$, $E_{\times}$ & The edge sets \\

    $W$, $W'$ & The edge weight sets \\

    $f(\cdot)$ & The weight mapping function \\

    $\mathscr{L}(\cdot)$, $\mathscr{F}(\cdot)$, $\mathscr{K}(\cdot)$ & The structural mapping functions\\

    $\mathcal{A}(\cdot)$ & The function for getting the attributed object \\

    $\mathcal{G}$ & The set of graphs \\
  
    $c$, $C$ & The label of the graph $G$ and the label set\\





    \hline\hline
  \end{tabular}
\end{table}

\subsection{Transaction Subgraph Networks} \label{subsec:defination}

In this section, we introduce the detail of our transaction subgraph network model. Firstly, we give the definitions of transaction subgraph network (TSGN), directed transaction subgraph network (Directed-TSGN), and temporal transaction network (Temporal-TSGN).
And then, we elaborate the construction methods of the TSGN, Directed-TSGN, and Temporal-TSGN, respectively.

\begin{definition}[\textbf{TSGN}]\label{def:1}
Given a transaction graph $G=(V,E)$ where $e=(v_k,v_j, w) \in E$ indicates a transaction edge with the weight $w$, the TSGN, denoted by $T=\mathscr{L}(G)$, is a mapping from $G$ to $T=(V',E')$, with the node and edge sets denoted by $V'=\{e_i \in E\}$ and $E'=\{(e_i,e_j,w')\}$, $i,j=0,1,2,\cdots$. The transaction subgraphs $e_a$ and $e_b$ will be connected if they share the common addresses in the transaction graph $G$.
\end{definition}

According to the Definition~\ref{def:1}, we can find that TSGN is the variant of SGN model~\cite{xuan2019subgraph} for the background of Ethereum transaction ecosystem. Different from SGN model, TSGN adds a network weight mapping mechanism  $W' \leftarrow f(W)$, which can retain the transaction amount information in the original transaction network for the downstream behaviour analysis task. The edge weights of TSGN can be calculated specifically by,
\begin{equation}
    w'=\left\{
   \begin{matrix}
   log(\frac{w_a+w_b}{2}), &  \text{if } w_a\ne 0 \text{ or } w_b \ne 0\\
    0,                     &  \text{if } w_a=0 \text{ and }w_b=0
  \end{matrix}
  \right.
  \label{eq1}
\end{equation}
where $w' \in W'$ is the mapped weights, and $w_a, w_b \in W$ are the original weights. Note that transaction amount is set to 0 when the account interaction relationship is the invocation of contract accounts. Here, TSGN is a classical model which can be applied for any simple Ethereum account interaction network. Considering the complex transaction direction information, we propose the variant named Directed-TSGN.

\begin{definition}[\textbf{Directed-TSGN}]\label{def:2}
Given a transaction digraph $G$ = $(V,E)$, where $d=(v_{src},v_{dst},w) \in E$ indicates a directed transaction edge, the Directed-TSGN, denoted by $T_D=\mathscr{F}(G)$, is a mapping from $G$ to $T_D=(V^*, E^*)$, with the node and edge sets denoted by $V^*=\{d_i \in E|i=0,1,2,\cdots\}$. There will be a directed interaction edge $(d_a, d_b, w^*) \in E^*$ between transaction subgraphs $d_a$ and $d_b$ when they meet the following conditions:
\begin{itemize}
\item transaction subgraphs $d_a$ and $d_b$ share the common addresses in the weighted directed transaction graph $G$;
\item the source of $d_a$ (or $d_b$) is the destination of $d_b$ (or $d_a$), which can form a 2-top path with the same direction.
\end{itemize}
\end{definition}

As shown in the Definition~\ref{def:2}, Directed-TSGN based on TSGN model introduces the direction information into the mapping mechanism which can capture the path of transaction behavior. The weight $w^*$ can be obtained by Eq. (\ref{eq1}). Also, Ethereum data is dynamic and gets larger and more complicated over time, which brings the challenge to analyze transaction behaviors by adopting some graph mining algorithms. In addition, phishing behaviors generally have the strong time-aware property. These challenges require the development of a novel graph model mapping approach that is suitable for the analysis of time-varying and dense networks. Hence, we further propose Temporal-TSGN.

\begin{figure}[t]
\centering
\includegraphics[width=\linewidth]{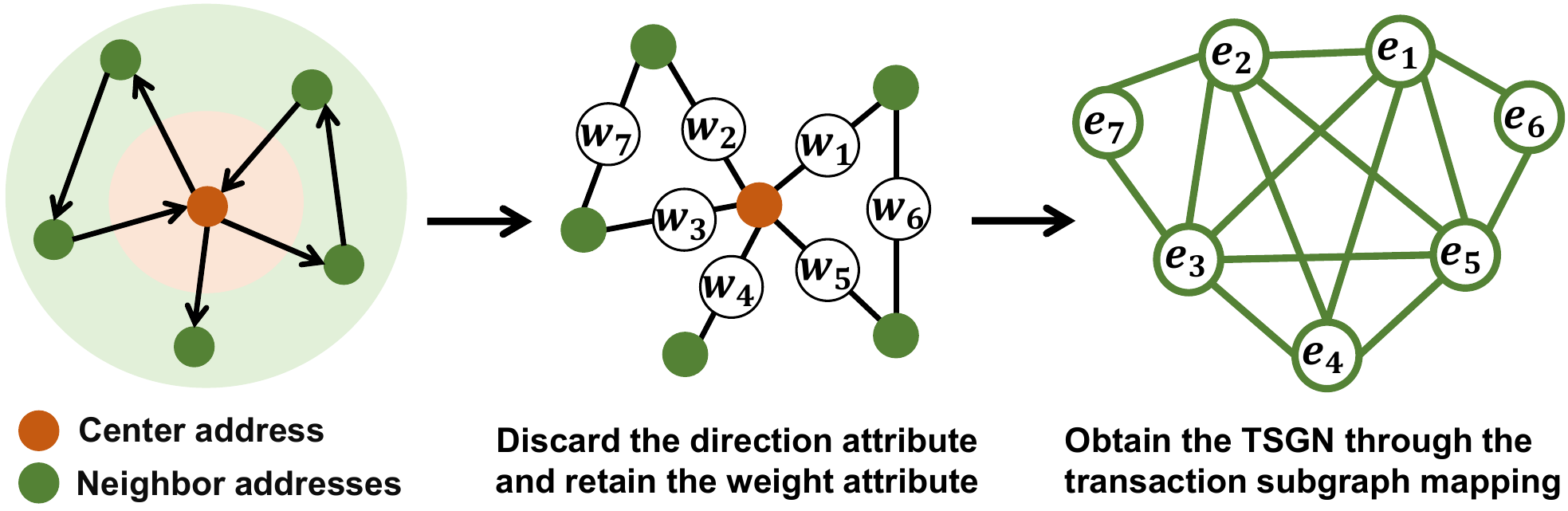}
\caption{The process of building TSGN from a given transaction network.}
\label{fig:TSGN}
\end{figure}

\begin{definition}[\textbf{Temporal-TSGN}]\label{def:3}
Given a temporal transaction digraph $G$ = $(V,E)$, where $t=(v_{src},v_{dst},w,\ell) \in E$ indicates a directed transaction edge with the timestamp $\ell$, the Temporal-TSGN, denoted by $T_T=\mathscr{K}(G)$, is a mapping from $G$ to $T_T=(V_{\times}, E_{\times})$, with the node and edge sets denoted by $V_{\times}=\{t_i \in E|i=0,1,2,\cdots\}$ and $E_{\times}\subseteq(V_{\times} \times V_{\times})$. A directed edge $(t_a,t_b,w_{\times}) \in E_{\times}$ will be built between two temporal directed transaction subgraphs $t_a$ and $t_b$ when they meet the stricter conditions: In the original temporal transaction graph $G$,
\begin{itemize}
\item $t_a$ and $t_b$ share the common addresses in the original directed transaction graph $G$;
\item the source of $t_a$ (or $t_b$) is the destination of $t_b$ (or $t_a$), which can form a 2-top path with the same direction.
\item the timestamp $\ell_a < \ell_b$ (or $\ell_b < \ell_a$) makes ($t_a,t_b$) (or ($t_a,t_b$)) present a sequential transaction flow.
\end{itemize}
\end{definition}

%
Clearly, the Temporal-TSGN defined in the Definition~\ref{def:3} is the substructure which is obtained by filtering the nodes and edges of of Directed-TSGN according to the time information, i.e., $T_T \subseteq T_D$. By attaching the time attribute on the directed transaction subgraphs (edges), Temporal-TSGN can capture neighbor transaction flow information for each target account address while reducing the scale of the network. The weight value $w_{\times}$ can also be calculated by Eq. (\ref{eq1}).
Next, we would focus on the specific construction methods for the TSGN, Directed-TSGN, and Temporal-TSGN.

\subsection{Construction of TSGN} \label{subsec:TSGN}

Fig.~\ref{fig:TSGN} shows the process of constructing TSGN. Given an original transaction network composed of a center address and its neighbor addresses, we can firstly get a plain transaction network with weight values by discarding the direction attribute and retaining the weight attribute. And then, this network is mapped into TSGN structural space according to the rule indicated in the Definition~\ref{def:1}.

Specifically, the edges in the undirected transaction network is mapping to the nodes of TSGN, and then new edges are built between nodes $e_1$, $e_2$, $e_3$, $e_4$, $e_5$ because the edges of undirected transaction network share the common center node. There is no doubt that transaction subgraph mapping strategy of TSGN also works for the case that there are connections between neighbor addresses. The transactions $e_6$ and $e_7$ are mapped into the TSGN to acquire the neighbor trading pattern information for the center address. A pseudocode of constructing TSGN is given in Algorithm \ref{alg:1}. The input of this algorithm is the original transaction network $G$($V$,$E$) and the output is the constructed TSGN, denoted by $T$($V'$,$E'$), where $V'$ and $E'$ represent the sets of nodes and edges in the TSGN, respectively. Here, we choose the weight mapping function introduced in Section~\ref{subsec:defination}. Of course, different weight mapping functions can be defined as required.

\begin{figure}[t]
\centering
\includegraphics[width=\linewidth]{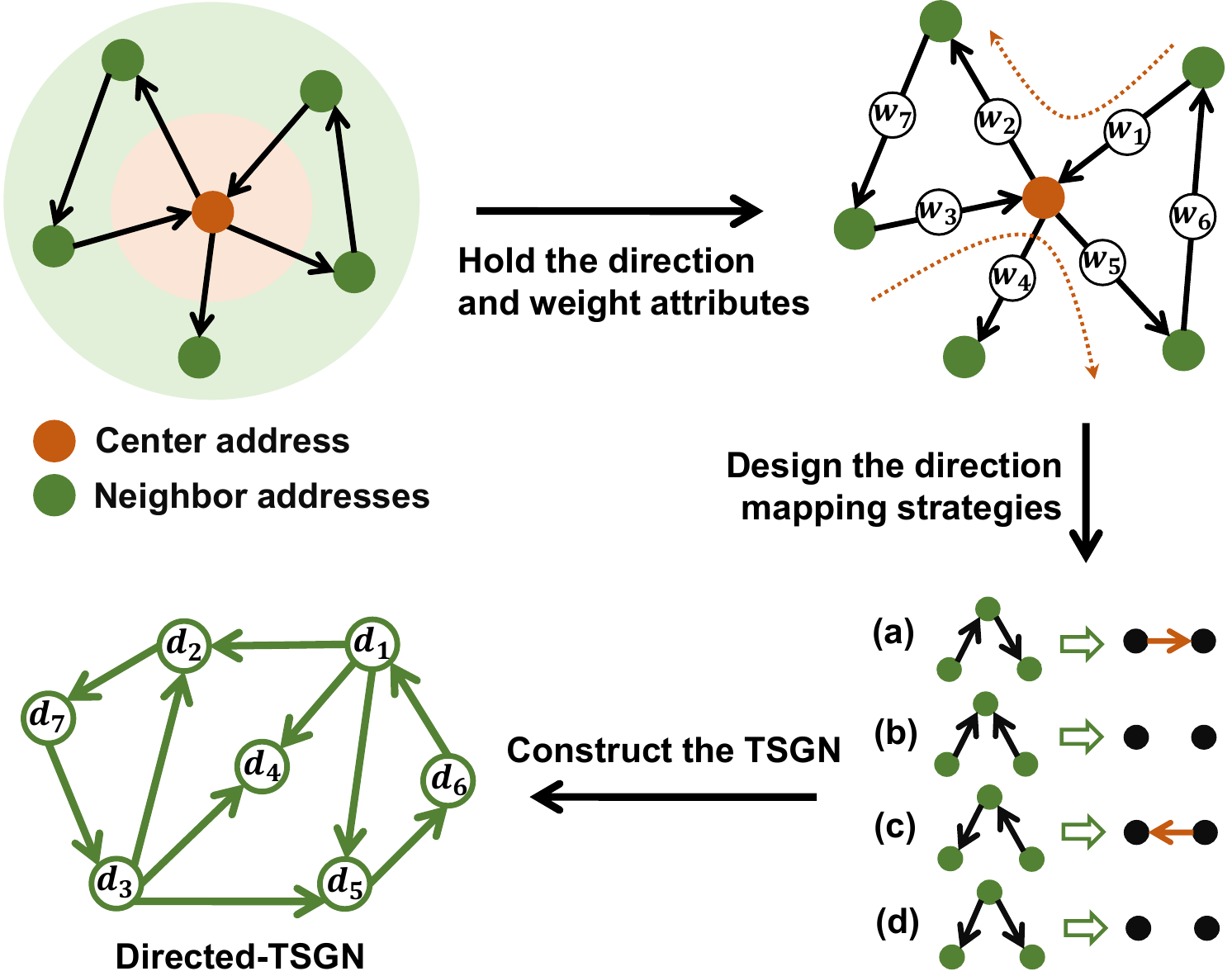}
\caption{The process of building Directed-TSGN from a given transaction network.}
\label{fig:D-TSGN}
\end{figure}

\subsection{Construction of Directed-TSGN} \label{subsec:D-TSGN}

Transaction direction is one of the most noteworthy characteristics in the field of financial research. Most of studies model the transaction pattern of an account by \emph{Indegree} and \emph{Outdegree} in a transaction graph. In Ethereum or Bitcion platform, transaction direction combined with transaction amount information can be adopted to study the peculiar unusual patterns~\cite{maesa2017detecting}, detect abnormal economic interactions~\cite{chen2020phishing}, predict virtual currency prices~\cite{abay2019chainnet} and so on. According to Section~\ref{subsec:TSGN}, we can find that the TSGN model can not retain the direction information which may play an important role in the following phishing identification tasks. Moreover, the TSGN is denser than the original transaction network and even becomes a fully connected network, which may reduce the representation ability of graph mining algorithms. Directed-TSGN proposed in this paper can take into account the two attributes of transaction direction and amount into consideration, benefiting for extracting the potential transaction patterns of the target addresses.

In response to the above problems, we propose one variant of TSGN, named Directed-TSGN. As shown in Fig.~\ref{fig:D-TSGN}, the construction of Directed-TSGN consists of the following three steps: (i) hold the direction and weight attributes, (ii) design the direction mapping strategies, (iii) construct the TSGN. The first step aims to hold the direction and weighted attributes of the original transaction network and prepare for the subsequent structural mapping. Similarly, the edges are mapped into the nodes $d_1$, $d_2$, $d_3$, $d_4$, $d_5$, $d_6$, $d_7$ of the Directed-TSGN. The two red directed dashed lines indicate that the transactions $d_1$ and $d_2$ and transactions $d_3$ and $d_5$ can be considered as two continuous transaction behaviors, respectively. 
In other words, the edges $d_1$ and $d_2$ and the edges $d_3$ and $d_5$ can form two 2-hop paths with the same direction, respectively. 
According to the four direction mapping strategies, we can build the new edges in the Directed-TSGN. Due to the fact that patterns of (b) and (d) don't satisfy the requirements of constructing edges, the Directed-TSGN can limit the network size and get a relatively sparse transaction subgraph network. The pseudocodes of constructing Directed-TSGN consist of Algorithm \ref{alg:1} and \ref{alg:2}. 
The input of Algorithm \ref{alg:2} is two transaction subgraphs $d_a,d_b$ in the attribute attached graph $G$, and the output is the edge $(d_a,d_b)$ matching the mapping rules. Algorithm \ref{alg:2} aiming to illustrate the Directed-TSGN mapping will be executed when $G \leftarrow \mathcal{A}_{directed}(G)$ is attached the transaction direction attribute.


\begin{figure}[t]
\centering
\includegraphics[width=\linewidth]{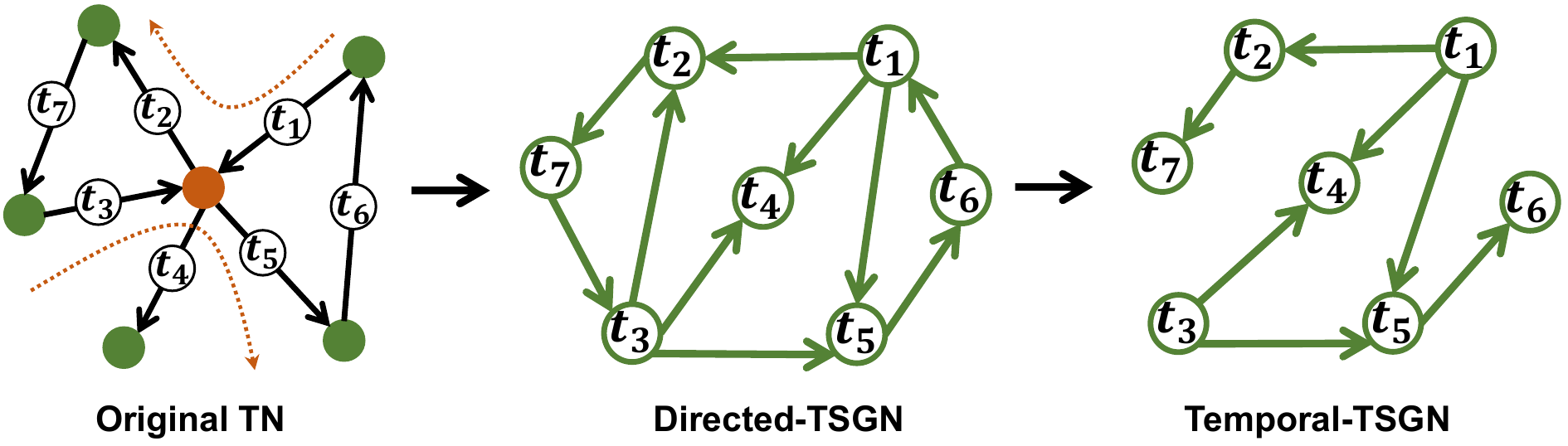}
\caption{The process of building Temporal-TSGN from a given original temporal transaction network.}
\label{fig:T-TSGN}
\end{figure}

\subsection{Construction of Temporal-TSGN} \label{subsec:T-TSGN}

Temporal information, i.e., transaction timestamps, can clearly indicate when the transaction occurred. Hence, the transaction time in alliance with trading flow direction offers the possibility for the traceability of transferring funds on Ethereum~\cite{lin2022ethereum}. In this section, we further introduce the temporal attribute of transactions into the TSGN mapping mechanism. On the basis of Directed-TSGN, Temporal-TSGN is proposed to fuse crucial temporal and direction attributes related to phishing detection for enhancing the ability to identify phishing scams. Temporal-TSGN improves the mapping strategies compared with Directed-TSGN, which can finely captures the temporal transaction flow structural information in Ethereum, further limiting the network scale.


Fig.~\ref{fig:T-TSGN} shows the progress of constructing Temporal-TSGN. Given an original transaction network, Directed-TSGN can be firstly obtained according to the strategy mentioned in Section~\ref{subsec:D-TSGN}, and then we can get the Temporal-TSGN from Directed-TSGN by filtering the edges that are not in chronological order. The timestamps on the edges determine the order in which transaction are generated, i.e., $(t_1,t_2,t_3,t_4,t_5,t_6,t_7)$.
We define that, in the Temporal-TSGN, the transaction $t_i$ must be linked to transaction $t_j$ where $i<j$, which can form a 2-hop sequential transaction flow. Here, the edges $(t_3,t_2)$, $(t_7,t_3)$, and $(t_6,t_1)$ are unsatisfied with the formation condition and are therefore ignored.

\begin{figure}[t]
\centering
\includegraphics[width=\linewidth]{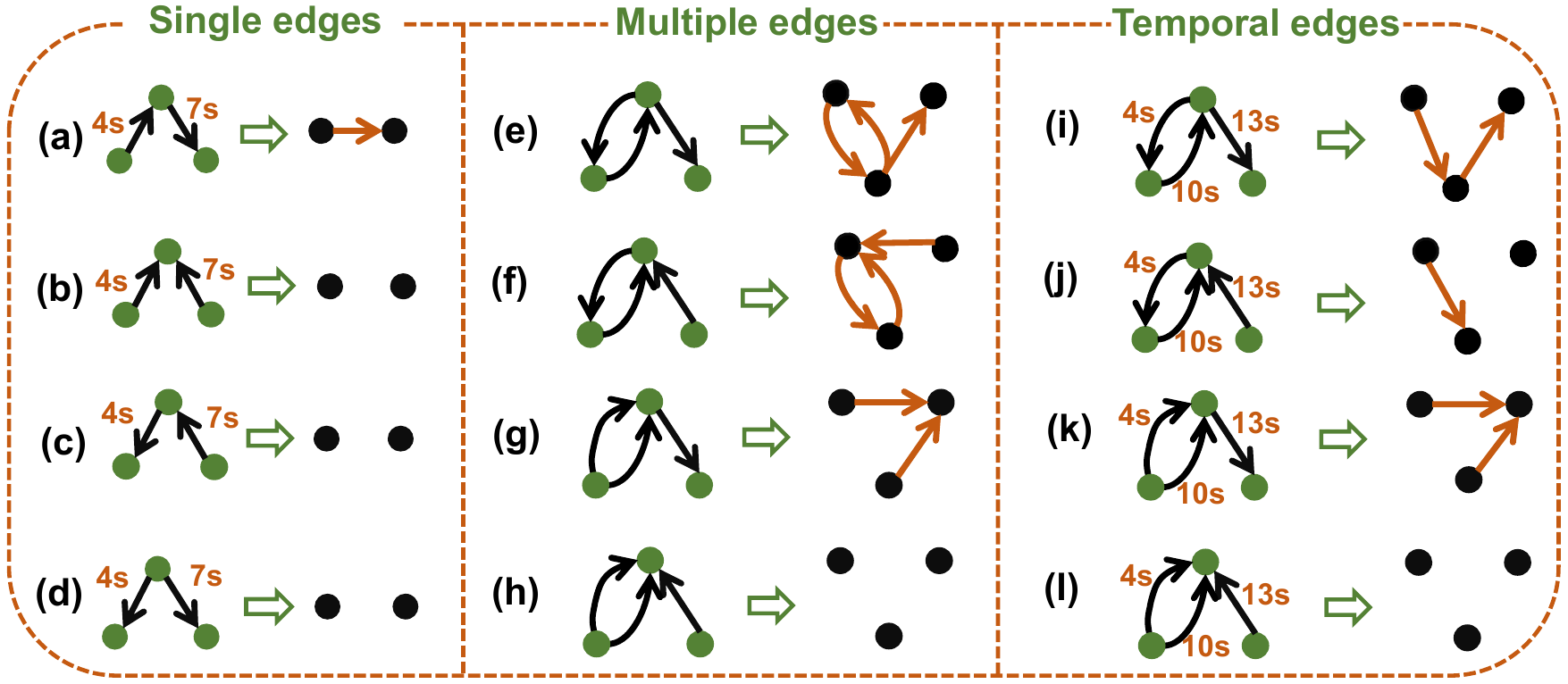}
\caption{Strategies of multiple edges and temporal edges cases for constructing TSGN.}
\label{fig:TD-TSGN}
\end{figure}

At this point, the Temporal-TSGN mapping strategy can be summarized as the \emph{single edges} part of Fig.~\ref{fig:TD-TSGN}. We attach the timestamps onto the edges of different directions, getting the fund flow of the transactions within a certain time frame. It is important to note that, a 2-hop transaction chain is available for the construction of Temporal-TSGN only under the three conditions described in the Definition~\ref{def:3}. Specifically, the transaction chain $(v_1\stackrel{4s}{\longrightarrow}v_2\stackrel{7s}{\longrightarrow}v_3)$ shown in Fig.~\ref{fig:TD-TSGN} (a) can be successfully mapped into the TSGN structural space, becoming a 1-hop directed node pair, while the other three transaction chains $(v_1\stackrel{4s}{\longrightarrow}v_2\stackrel{7s}{\longleftarrow}v_3)$, $(v_1\stackrel{4s}{\longleftarrow}v_2\stackrel{7s}{\longleftarrow}v_3)$, and $(v_1\stackrel{4s}{\longleftarrow}v_2\stackrel{7s}{\longrightarrow}v_3)$ shall be excluded since they do not satisfy the second condition or the third condition or both in the Definition~\ref{def:3}, respectively.

%
%
%
%

\begin{algorithm}[t]
\caption{\textbf{Constructing TSGN.}}
\label{alg:1}
\KwIn{A transaction network $G$($V$,$E$) with transaction addresses set $V$, transaction edges set $E\subseteq(V\times{V})$.}
\KwOut {TSGN with the attribute, denoted by $T_{\mathcal{A}}$.} 
Initialize $T_{\mathcal{A}}$ with node set $V'$ and edge set $E'$\;
\For {each $v \in V$}
{
    Get the neighbor addresses set $\mathcal{N}(v)$\;
      \For {each $u \in \mathcal{N}(v)$}
          {$ txn \leftarrow $ sorted([$v$, $u$])\;
          Add $txn$ into the set $trans_{pool}$\;
          }
      \For {$txn_a,txn_b \in trans_{pool}$ and $txn_a \neq txn_b$}
          {$G \leftarrow \mathcal{A}(G)$\; 
          \tcp{{\scriptsize attach the transaction attributes to G}}
          \If {$G$ is directed}
          {$d_a,d_b \leftarrow \mathcal{A}(txn_a),\mathcal{A}(txn_b)$\;
          Get ($d_a,d_b$) by executing Algorithm \ref{alg:2}\;
          \tcp{{\scriptsize Directed-TSGN mapping}}
          Add the edge ($d_a,d_b$) into $E'$\;}
          \If {$G$ is directed and time-aware}
          {$t_a,t_b \leftarrow \mathcal{A}(txn_a),\mathcal{A}(txn_b)$\;
          Get ($t_a,t_b$) by executing Algorithm \ref{alg:3}\; 
          \tcp{{\scriptsize Temporal-TSGN mapping}}
          Add the edge ($t_a,t_b$) into $E'$\;}
          \Else{Add ($txn_a,txn_b$) into $E'$\;} 
          }
    Add $trans_{pool}$ into $V'$;
}
\Return $T_{\mathcal{A}}$($V'$,$E'$) with corresponding attribute\;
\end{algorithm}

\begin{algorithm}[!t]
\caption{\textbf{Directed-TSGN mapping.}}
\label{alg:2}
\KwIn{Directed $d_a, d_b \in trans_{pool}$.} 
\KwOut {Directed edge ($d_a,d_b$).} 
\If {dst($d_a$) = src($d_b$)}{Add a directed edge from $d_a$ to $d_b$;}
\If {src($d_a$) = dst($d_b$)}{Add a directed edge from $d_b$ to $d_a$;}
\Else {continue;}
\Return ($d_a,d_b$) with direction attribute\;
\end{algorithm}

\begin{algorithm}[!t]
\caption{\textbf{Temporal-TSGN mapping.}}
\label{alg:3}
\KwIn{Directed temporal $t_a, t_b \in trans_{pool}$.} 
\KwOut {Directed edge ($t_a,t_b$).} 
\If {dst($t_a$) = src($t_b$)}{\If {timestamp($t_a$) $<$ timestamp($t_b$)}{Add a directed edge from $t_a$ to $t_b$;}}
\If {src($t_a$) = dst($t_b$)}{\If {timestamp($t_a$) $>$ timestamp($t_b$)}{Add a directed edge from $t_b$ to $t_a$;}}
\Else {continue;}
\Return ($t_a,t_b$) with direction attribute\;
\end{algorithm}

\begin{figure}[h]
\centering
\includegraphics[width=\linewidth]{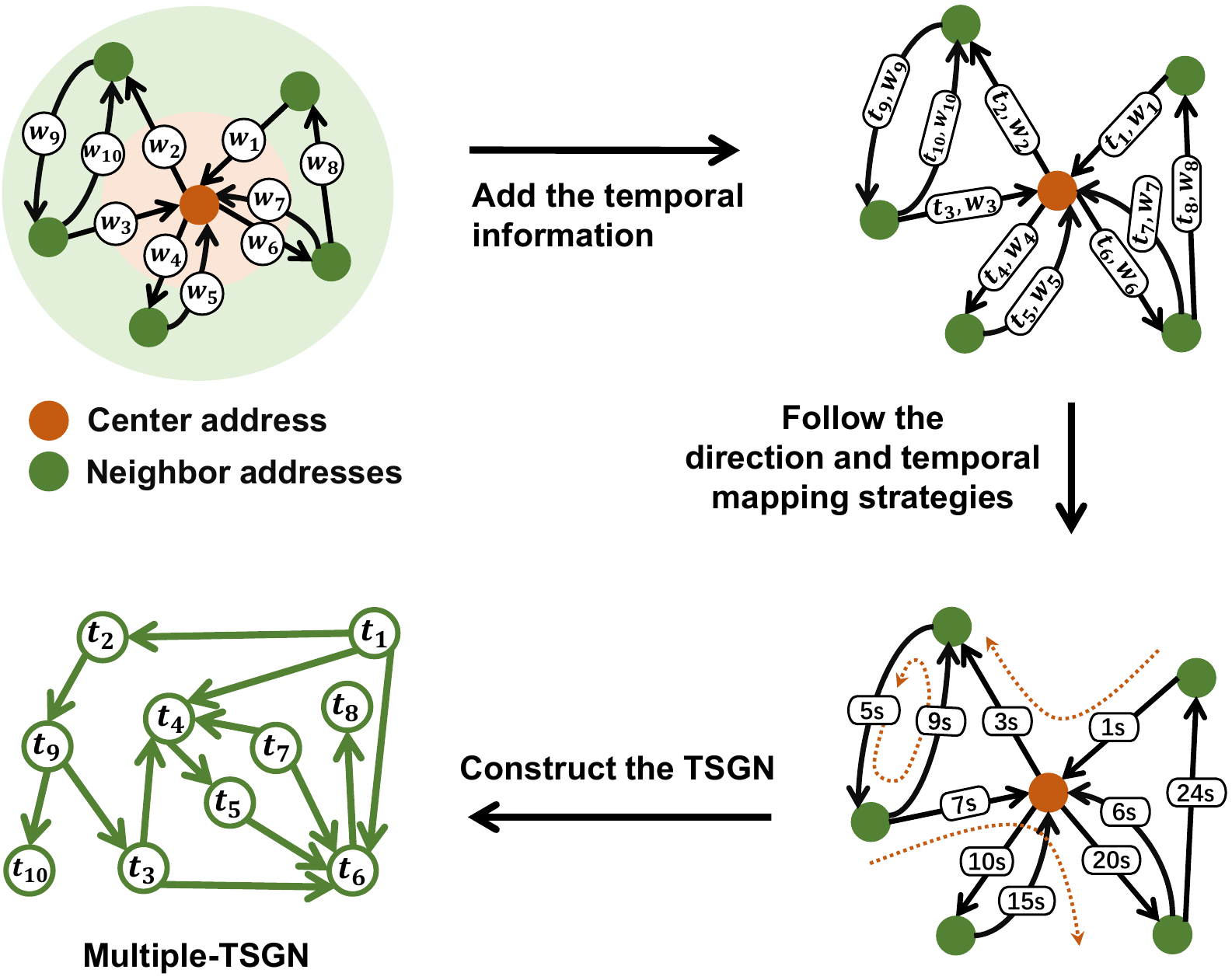}
\caption{The progress of building TSGN for the multi-edge case and obtain Multiple-TSGN.}
\label{fig:M-TSGN}
\end{figure}

\subsection{TSGN for Multi-edge Case} \label{subsec:M-TSGN}

Ethereum transaction data poses following several challenges. First, the transaction network is generally sparse and dynamic, but it chould be very dense when viewed over a long period of time.
Secondly, the nodes in the transaction graph appear (i.e., new accounts are created) and disappear (i.e., no future transactions occur) daily.
Thirdly, transaction throughput on a daily basis fluctuates greatly.
So, the time attribute is a vitally important factor when constructing transaction networks, which can exhibit the real complexity of the Ethereum transaction network from the time dimension. Li~\cite{li2022ttagn} modeled the temporal relationship of historical transaction records between nodes to construct the edge representation of the Ethereum transaction network. In this section, how to deal with the mapping rules of multi-edge temporal networks is an urgent problem to be solved for the subsequent phishing behavior analysis.

It is worth to note that the temporal mapping mechanism can be extended to the case of multiple edges. As shown in Fig.~\ref{fig:TD-TSGN} (e)-(l), the case of multi-edge mapping does not alter the loop of a nodes pair (Fig.~\ref{fig:TD-TSGN} (e) and (f)), while the case of temporal edges can tackle it (Fig.~\ref{fig:TD-TSGN} (i) and (j)). The timestamps can break the cycle in the network structure and orient the information flow in the TSGN. Intuitively, TSGN under multi-edge case can yield to a clear path of funding transferring and capture the fund flow pattern in the complex temporal transaction network, which is conducive to the detection of illegal trading activities.

In particular, the construction progress of TSGN for multi-edge transaction network, denoted as Multiple-TSGN, is shown in Fig.~\ref{fig:M-TSGN}. Firstly, given a weighted directed multi-edge graph, it will turn to a MultiDiGraph where each transaction has its own fixed order of occurrence after being attached with the timestamp attribute. And then, according to the Algorithm~\ref{alg:3}, we can get the variant of Temporal-TSGN for multi-edge transaction network, i.e., Multiple-TSGN.

To date, it is still the research focus to deeply understand and analyze the knowledge of multi-edge temporal transaction networks in Ethereum.
In this section, Multiple-TSGN proposed provides a novel perspective to excavate the potential transaction patterns in Ethereum.

\begin{table*}[t]
\caption{Basic statistics of six Ethereum datasets after adding different transaction attributes. \emph{Type} denotes the type of network structure, $N_G$ is the number of graphs, \#$C_{max}$ is the number of graphs belonging to the largest class, $N_C$ is the number of classes, \#N and $max$\#N are the average and maximum numbers of the nodes, and \#E and $max$\#E are the average and maximum numbers of edges, respectively, of the graphs in the dataset. Note that, we display three different statistics for each dataset, such as \emph{Plain}, \emph{Directed}, and \emph{Multiedge}.}
\label{tab-data}
\centering
\renewcommand{\arraystretch}{1.5}
\setlength{\tabcolsep}{3.1mm}{
\begin{tabular}{cccccrrrrrrrr}
\hline\hline
\multirow{2}*{Dataset}&\multirow{2}*{Type}&\multirow{2}*{$N_G$}&\multirow{2}*{\#$C_{max}$}&\multirow{2}*{$N_C$}&\multirow{2}*{\#N}&\multirow{2}*{$max$\#N}
&\multicolumn{2}{c}{Plain}&\multicolumn{2}{c}{Directed}&\multicolumn{2}{c}{Multiedge}  \\ \cline{8-13}
   &  &  &  &  &  &  & \#E    & $max$\#E    & \#E    & $max$\#E    & \#E    & $max$\#E    \\
\hline
EtherG1 & Star & 700 & 350 & 2 & 7 & 13 & 6 & 12 & 6 & 12 & 8 &13 \\
EtherG2 & Star & 700 & 350 & 2 & 14 & 33 & 13 & 32 & 13 & 32 & 20 &34 \\
EtherG3 & Star & 700 & 350 & 2 & 96 & 4972 & 95 & 4972 & 100 & 4985 & 23 &12209 \\
EtherG4 & Net  & 700 & 350 & 2 & 7 & 13 & 7 & 20 & 7 & 21 & 163 &24317 \\
EtherG5 & Net  & 700 & 350 & 2 & 14 & 33 & 17 & 58 & 18 & 65 & 465 &24354 \\
EtherG6 & Net  & 700 & 350 & 2 & 96 & 4972 & 161 & 7534 & 174 & 10745 & 1765 &68959 \\
\hline\hline
\end{tabular}
}
\end{table*}

\section{Experiments}\label{sec:experiment}


In this section, we describe the experiments conducted to evaluate the effectiveness of our TSGN models on the task of phishing detection. We first construct the network datasets by exploiting transactions records from Ethereum, then outline several baselines followed by the corresponding experimental setup. Finally, we show the experimental results with discussion.


\subsection{Data Collection and Preparation}\label{subsec:data}


As today's largest blockchain-based application, Ethereum has fully open transaction data which are permanently recorded and easily accessed through the API of Etherscan~\footnote{\url{https://etherscan.io}} or the academic blockchain data platform XBlock~\footnote{\url{http://xblock.pro/}}.

Ethereum has gained massive momentum in recent years, mainly due to the success of cryptocurrencies.
Considering that the entire cryptocurrency trading network is enormous, we crawl some phishing addresses and non-phishing addresses as the target nodes and extract their first-order neighbor nodes from the first seven million blocks of Ethereum to construct transaction network datasets (all transactions occerred between July 30, 2015 and January 2, 2019).
After filtering and preprocessing the raw data, we randomly select 1050 phishing addresses and 1050 unlabeled addresses according based on the account label information on the XBlock platform, and crawl their neighbor addresses and transaction records to construct transaction networks. And then, these networks will be randomly divided so that we finally get three balanced \emph{star-form} network datasets EtherG1, EtherG2, and EtherG3, each of which has 350 transaction networks of phishing addresses and 350 transaction networks of non-phishing addresses. Besides, in order to increase the structural complexity of the transaction networks, we expand the scale of the networks by introducing the interactive edges between 1-hop neighbors, yielding to three \emph{net-form} transaction datasets EtherG4, EtherG5, and EtherG6. The basic statistics of these datasets are presented in TABLE~\ref{tab-data}.

\begin{figure}[h]
\centering
\includegraphics[width=\linewidth]{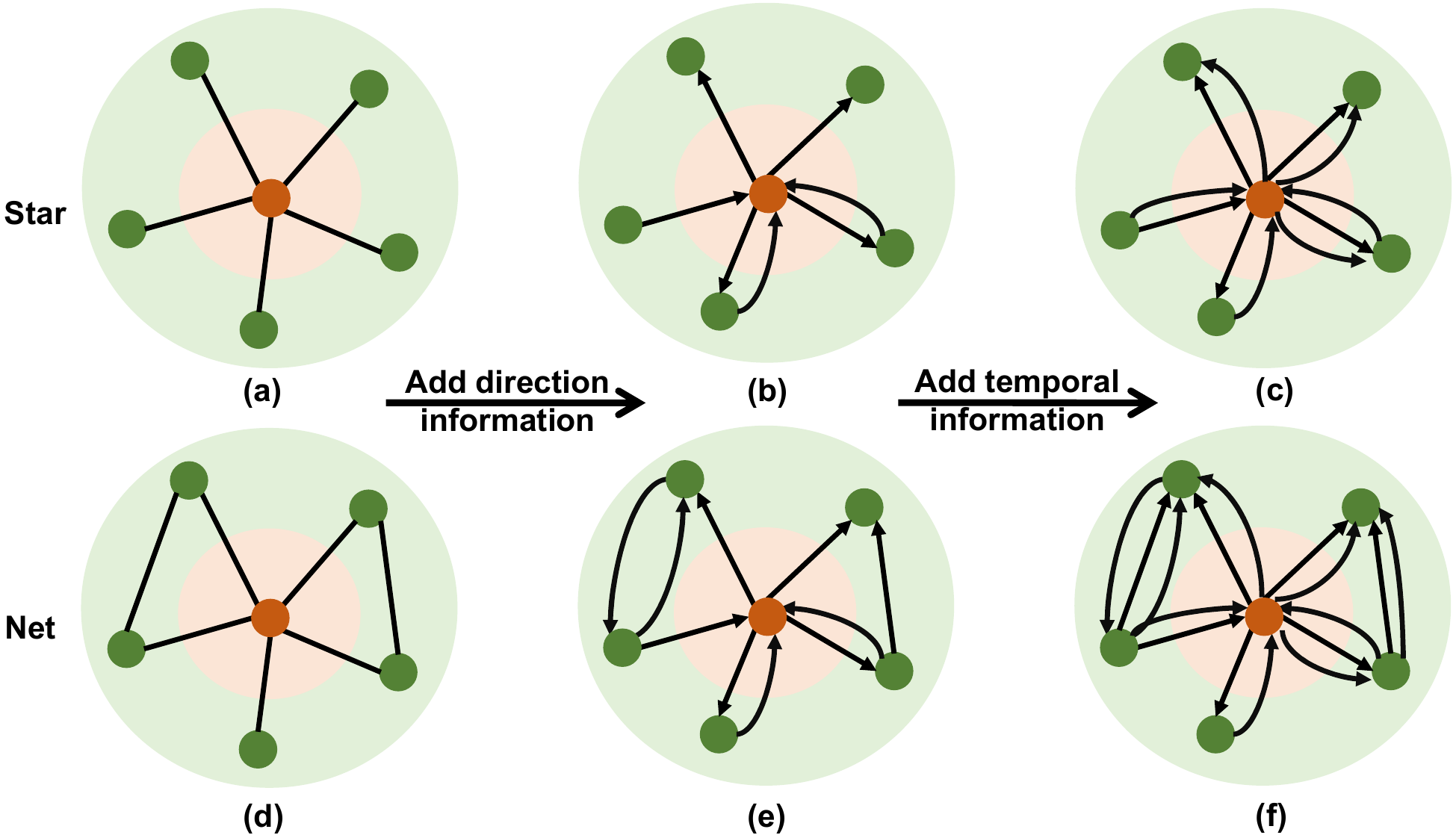}
\caption{Illustration of datasets where network structures evolving with the attributes.}
\label{fig:datasets}
\end{figure}

As shown in TABLE~\ref{tab-attr}, different variants of TSGN model have their own unique attribute dependencies. 
Hence, it is necessary to sketch the transaction network structure under different attributes. 
Based on Fig.~\ref{fig:datasets}, we further illustrate the details of edge statistics of TABLE~\ref{tab-data}.
\emph{Plain} indicates no directed and temporal attributes are included in the network structures (Fig.~\ref{fig:datasets} (a) and (d)), \emph{Directed} means the edges of networks are directed (Fig.~\ref{fig:datasets} (b) and (e)), and \emph{Multiedge} signifies the edges are attached directed and temporal attributes simultaneously to distinguish the multiple same-directed edges between a node pair (Fig.~\ref{fig:datasets} (c) and (f)). Note that, the weight is set as the inherent attribute. Apparently, the richer the edge attributes, the more complex the network structures. And the next experiments will be conducted on these six datasets to verify the performance of the four TSGN models.

\begin{table}[ht]
\caption{Network attribute statistics required by different TSGN models.}
\label{tab-attr}
\centering
\renewcommand{\arraystretch}{1.5}
\setlength{\tabcolsep}{1.5mm}{
\begin{tabular}{lcccc}
\hline\hline
Model & Weighted & Directed & Temporal & Multiedge \\
\hline
TSGN & \Checkmark & \XSolidBrush & \XSolidBrush & \XSolidBrush \\
Directed-TSGN & \Checkmark & \Checkmark & \XSolidBrush & \XSolidBrush\\
Temporal-TSGN & \Checkmark & \Checkmark & \Checkmark & \XSolidBrush\\
Multiple-TSGN & \Checkmark & \Checkmark & \Checkmark & \Checkmark\\
\hline\hline
\end{tabular}
}
\end{table}

\subsection{Baseline Methods}\label{subsec:graphrepre}


In order to better verify the effect of the proposed TSGN models, we adopt six Ethereum phishing scams detection models, namely handcrafted features, Graph2Vec, DeepKernel, Diffpool, U2GNN and Line\_Graph2Vec which are introduced in the following.

\textit{Handcrafted Features}:
Handcrafted features are the most straightforward method to depict structure properties of networks. Different from previous feature designs listing a lot transaction attributes, we adopted ten classic topological attributes in network science which are widely used in graph mining tasks~\cite{li2011graph,xuan2019subgraph,wang2021sampling,fu2018link}. As a baseline, we aim at building the representation of the transaction networks by manually extracting the features for the subsequent account identification task. The first six handcrafted features are the basic statistics of topological structures, including \emph{Number of addresses}, \emph{Number of transactions}, \emph{Average degree of transaction subgraph}, \emph{Percentage of leaf nodes}, \emph{Network density}, and \emph{Average neighbor degree}. And the last four features, mainly including \emph{Average clustering coefficient}, \emph{Largest eigenvalue of the adjacency matrix}, \emph{Average betweenness centrality}, and \emph{Average closeness centrality}, are the relatively complex network computing properties. Here, \emph{Average clustering coefficient} is a very popular metric to quantify the edge density in ego networks. \emph{Eigenvalues} as the isomorphic invariant of a graph can be used to estimate many static attributes, such as connectivity, diameter, of transaction networks. \emph{Centrality} measures are indicators of the importance (status, prestige, standing, and the like) of a node in a network. Above various features provide a comprehensive portrait of the transaction network.


\textit{Graph2Vec}~\cite{Narayanan2016graph2vec}: It is the first unsupervised embedding approach for the entire networks, which constructs Weisfeiler-Lehman tree features for nodes in graphs and then learn graph representations by decomposing the graph-feature co-occurrence matrix.


\textit{DeepKernel}~\cite{yanardag2015deep}: This method provides a unified framework, named DGK, that defines the similarity between graphs via leveraging the dependency information of sub-structures, while learning latent representations. 


\textit{Diffpool}~\cite{ying2018hierarchical}: This method proposed a differentiable graph pooling module, which can generate hierarchical representations of graphs and can be combined with various GNN architectures in an end-to-end manner. This method mainly solves the problem that the traditional GNN methods are flat and can't learn the hierarchical representations of graphs.

\textit{U2GNN}~\cite{NguyenUGformer}: It presented a GNN model leveraging a transformer self-attention mechanism followed by a recurrent transition to induce a powerful aggregation function to learn graph representations, which achieved state-of-the-art performance on many classical benchmark for graph classification.

\textit{Line\_Graph2Vec}~\cite{yuan2020phishing}: This method collected a set of subgraphs, each of which contains a target address and its surrounding transaction network, and then converted them into the corresponding line graphs which can be input into the Graph2Vec model for acquiring the representation vectors for all the graphs.

\begin{table*}[!t]
\caption{Classification performance of our TSGN models on six Ethereum datasets, combining different graph classification methods.}
\label{tab-result}
\centering
\renewcommand{\arraystretch}{1.5}
\setlength{\tabcolsep}{1.0mm}{
\begin{tabular}{cllccccccc}
\hline\hline
& \multirow{3}*{\textbf{Method}} & \multirow{3}*{\textbf{Dataset}} & \multicolumn{7}{c}{\textbf{Classification results} ($F1$-$Score$, $\%$)} \\ \cline{4-10}
& & & \multicolumn{3}{c}{Star} & \multicolumn{3}{c}{Net} & \multirow{2}*{Avg.}\\
\cline{4-9}
& & & EtherG1 & EtherG2 & \multicolumn{1}{c}{EtherG3} & EtherG4 & EtherG5 & EtherG6 &  \\
\hline
\multirow{6}*{Baseline}
& Feature-based & Handcrafted  &  $74.29\pm{4.40}$      & $79.86\pm{3.01}$      & $83.86\pm{2.64}$      &  $76.21\pm{2.39}$      & $81.35\pm{3.48}$      & $86.64\pm{2.17}$      &80.36\\
\cdashline{2-10}[1pt/1pt]
& \multirow{2}*{Embedding} & 
Graph2Vec  &  $69.07\pm{2.95}$      & $71.79\pm{3.31}$      & $64.43\pm{2.51}$      &  $72.57\pm{3.29}$      & $73.00\pm{4.44}$      & $70.86\pm{2.59}$      &70.29\\
& & DeepKernel  &  $72.71\pm{5.09}$      & $75.29\pm{3.99}$      & $73.57\pm{3.35}$       &  ${\bf76.43\pm{5.08}}$ & $73.71\pm{5.50}$      & $74.57\pm{4.03}$      &74.38\\
\cdashline{2-10}[1pt/1pt]
& \multirow{2}*{GNN-based} & 
Diffpool  &  $81.29\pm{3.47}$      & $87.00\pm{1.74}$      & $95.00\pm{2.49}$       &  $93.36\pm{1.20}$      & $92.36\pm{2.30}$      & $93.79\pm{1.36}$      &90.47\\
& & U2GNN  &  $76.57\pm{3.45}$ &      $76.57\pm{4.99}$ &      $70.43\pm{5.57}$ &      $75.85\pm{4.67}$ &      $75.71\pm{5.31}$ &      $73.00\pm{3.03}$       &74.69 \\
\cdashline{2-10}[1pt/1pt]
& \multicolumn{2}{c}{Line\_Graph2Vec}          & $69.57\pm{3.32}$      & $72.43\pm{4.65}$      & $70.29\pm{6.11}$     &  $73.29\pm{3.56}$      & $73.00\pm{3.47}$      & $72.57\pm{4.37}$      &71.86\\
\hline

\multirow{20}*{Ours} 
& \multicolumn{2}{l}{Handcrafted w/ TSGN}          &  $75.57\pm{3.35}$      & $79.00\pm{3.44}$      & $85.43\pm{3.49}$      &  $77.86\pm{5.47}$      & $81.86\pm{3.89}$      & $88.29\pm{3.71}$      &81.34\\
& \multicolumn{2}{l}{Handcrafted w/ Directed-TSGN} &  $77.50\pm{5.02}$      & $80.07\pm{3.42}$      & $85.00\pm{2.21}$      &  $81.07\pm{2.46}$      & $83.35\pm{3.13}$      & ${\bf89.57\pm{1.87}}$ &82.76\\
& \multicolumn{2}{l}{Handcrafted w/ Temporal-TSGN} &  ${\bf84.14\pm{2.51}}$ & ${\bf85.71\pm{4.65}}$ & ${\bf88.86\pm{4.18}}$ &  ${\bf81.43\pm{4.24}}$ & ${\bf85.71\pm{3.94}}$ & $89.43\pm{3.45}$      &{\bf85.88}\\
& \multicolumn{2}{l}{$\%Increase$} &13.26\%   &7.32\%   &5.96\%    &  6.85\%  &  5.36\%   & 3.38\%   & 6.87\% \\
\cdashline{2-10}[1pt/1pt]
& \multicolumn{2}{l}{Graph2Vec w/ TSGN}           & $69.57\pm{3.32}$      & $72.43\pm{4.65}$      & $70.29\pm{6.11}$     &  $73.29\pm{3.56}$      & $73.00\pm{3.47}$      & $72.57\pm{4.37}$      &71.86\\
& \multicolumn{2}{l}{Graph2Vec w/ Directed-TSGN} &  $73.57\pm{4.05}$      & $73.00\pm{3.44}$      & $68.50\pm{3.30}$      &  $73.57\pm{4.00}$      & $76.79\pm{2.81}$      & $71.50\pm{3.94}$      &72.82\\
& \multicolumn{2}{l}{Graph2Vec w/ Temporal-TSGN} &  ${\bf76.43\pm{3.57}}$ & ${\bf79.00\pm{5.15}}$ & ${\bf73.43\pm{2.57}}$ &  ${\bf76.00\pm{5.60}}$ & ${\bf79.29\pm{4.88}}$ & ${\bf75.57\pm{5.55}}$ &{\bf76.62}\\
& \multicolumn{2}{l}{$\%Increase$} &10.66\%   &10.04\%   &13.97\%    &  4.73\%  &  8.62\%   & 6.65\%   & 9.01\% \\
\cdashline{2-10}[1pt/1pt]

& \multicolumn{2}{l}{DeepKernel w/ TSGN}          &  $74.29\pm{5.61}$      & $77.57\pm{6.13}$      & $68.71\pm{5.73}$       &  $75.29\pm{5.68}$      & $75.29\pm{2.65}$      & $75.86\pm{3.53}$      &74.50\\
& \multicolumn{2}{l}{DeepKernel w/ Directed-TSGN} &  $73.86\pm{2.64}$      & $73.86\pm{4.19}$      & $69.71\pm{4.94}$       &  $73.86\pm{3.56}$      & $77.14\pm{6.48}$      & ${\bf78.00\pm{4.00}}$ &74.41\\
& \multicolumn{2}{l}{DeepKernel w/ Temporal-TSGN} &  ${\bf77.71\pm{3.63}}$ & ${\bf78.57\pm{4.61}}$ & ${\bf75.00\pm{4.39}}$  &  $76.14\pm{4.91}$      & ${\bf78.71\pm{4.58}}$ & $77.57\pm{5.82}$      &{\bf77.26}\\
& \multicolumn{2}{l}{$\%Increase$} &6.88\%   &4.36\%   &1.94\%    &  -0.38\%  &  6.78\%   & 4.60\%   & 3.87\% \\
\cdashline{2-10}[1pt/1pt]
& \multicolumn{2}{l}{Diffpool w/ TSGN}          &  $89.71\pm{3.05}$      & ${\bf96.29\pm{1.83}}$ & $96.29\pm{2.65}$       &  $91.43\pm{3.00}$      & ${\bf96.57\pm{1.59}}$ & $95.57\pm{1.49}$      &94.31\\
& \multicolumn{2}{l}{Diffpool w/ Directed-TSGN} &  ${\bf93.14\pm{1.78}}$      & $92.29\pm{2.73}$      & ${\bf97.00\pm{1.35}}$       &  $93.43\pm{1.10}$      & $94.07\pm{1.36}$      & $94.29\pm{1.84}$      &94.03\\
& \multicolumn{2}{l}{Diffpool w/ Temporal-TSGN} &  $89.29\pm{2.95}$ & $94.71\pm{2.12}$      & $96.71\pm{1.81}$  &  ${\bf94.14\pm{3.16}}$ & $96.14\pm{2.56}$      & ${\bf98.43\pm{1.35}}$ &{\bf94.90}\\
& \multicolumn{2}{l}{$\%Increase$} &14.58\%   &10.68\%   &2.10\%    &  0.84\%  &  4.56\%   & 4.95\%   & 4.90\% \\
\cdashline{2-10}[1pt/1pt]
& \multicolumn{2}{l}{U2GNN w/ TSGN}          &  $75.29\pm{4.09}$ &      $76.00\pm{4.77}$ &      $71.29\pm{4.58}$  &     $75.29\pm{5.11}$ &      $76.00\pm{3.92}$ &      $77.14\pm{2.47}$       &75.17 \\
& \multicolumn{2}{l}{U2GNN w/ Directed-TSGN} &  $76.00\pm{4.03}$ &      $77.71\pm{3.84}$ &      $72.86\pm{3.78}$  &     $76.43\pm{3.63}$ &      $78.57\pm{3.50}$ &      ${\bf78.00\pm{3.63}}$  &76.60 \\
& \multicolumn{2}{l}{U2GNN w/ Temporal-TSGN} &  ${\bf80.71\pm{3.27}}$ & ${\bf84.71\pm{3.56}}$ & ${\bf73.29\pm{4.47}}$ & ${\bf79.14\pm{4.66}}$ & ${\bf82.00\pm{2.80}}$ & ${\bf78.00\pm{3.79}}$  &{\bf79.64} \\
& \multicolumn{2}{l}{$\%Increase$} &5.41\%   &10.63\%   &4.06\%    &  4.34\%  &  8.31\%   & 6.85\%   & 6.63\% \\
\hline\hline
\end{tabular}}
\end{table*}


\subsection{Experimental Setup}\label{subsec:Expsetup}

To investigate the performance of our TSGN models, 
above mentioned representation methods are adopted to extract the feature vectors for TN, TSGN, Directed-TSGN, and Temporal-TSGN, preparing for training the classification model.
Here, the parameters setting of representation models under different TSGNs are given as follows.

For \textit{Handcrafted Features}, there are no hyperparameters, just extract 10 features manually for each graph as the feature vectors to train the phishing detection model.

For \textit{Graph2Vec} and \textit{Line\_Graph2Vec}, the parameter height of the WL kernel is set to 3. The commonly-used value of 1,024 is adopted as the embedding dimension. The other parameters are set to defaults: the learning rate is set to 0.025 and the epoch is set to 1000.

For \textit{DeepKernel}, the Weisfelier-Lehman subtree kernel is used to build the corpus and its height is set to 2, the embedding dimension is set to 10, the window size is set to 5 and skip-gram is used for the word2vec model. Furthermore, the node degrees are set as the node labels for TN and all TSGNs. As the key point of this method, the sub-structure similarity matrix $\mathcal{M}$, is calculated by the matrix $\mathcal{V}$ with each column representing a sub-structure vector, $\mathcal{M}=\mathcal{V}\mathcal{V}^\mathrm{T}$. Let $\mathcal{P}$ denote the matrix with each column representing a sub-structure frequency vector. According to the definition of kernel: $\mathcal{K} = \mathcal{P}\mathcal{M}\mathcal{P}^\mathrm{T} = \mathcal{P}\mathcal{V}\mathcal{V}^\mathrm{T}\mathcal{P}^\mathrm{T}=\mathcal{H}\mathcal{H}^\mathrm{T}$, one can use the columns in the matrix $\mathcal{H}=\mathcal{P}\mathcal{V}$ as the inputs to the classifier.

For \textit{Diffpool}, we use the in-degree and out-degree as the node features for the TN, and initialize the feature matrix by the edge weights of TN for different TSGNs. The other parameters of the model are set to defaults.

For \textit{U2GNN}, we use the in-degree and out-degree as the node features for the TN, and initialize the feature matrix by the edge weights of TN for different TSGNs. The other parameters of the model are set to defaults.

Next, for each feature representation method, the feature vectors of TN is respectively concatenated with those of different TSGNs, and then the dimension of the feature vectors is reduced to the same as that of the feature vector obtained from the original transaction network using PCA in the experiments, for a fair comparison.
For each dataset, one randomly split it into 9 folds for training and 1 fold for testing.
In order to accurately evaluate the quality of each classification model, in this paper, we will use \textit{$F_1$-$Score$} as a metric,
\begin{equation}
    F_1=\frac{2PR}{P+R},
\end{equation}
where \textit{P} is precision and \textit{R} is recall. \textit{$F_1$-$Score$} is the harmonic mean of precision and recall, so it can more comprehensively judge the pros and cons of the classification models.  To exclude the random effect of fold assignment, the experiment is repeated 300 times using the \emph{Random Forest} classifier and then the average $F_1$-$Score$ and its standard deviation recorded. Moreover, \textit{Percentage Increase} ($\%Increase$) is calculated to measure the performance improvement of the proposed models relative to the original model.
\begin{equation}
    \%Increase =\frac{F_1^{model}-F_1^{ori}}{F_1^{ori}}\times 100\%,
\end{equation}

\subsection{Experimental Results and Analysis}\label{subsec:result}

According to the above setting, we conduct some experiments on the six Ethereum datasets. To investigate the effectiveness of TSGN models, for each feature extraction algorithm, the classification results are compared on the basis of TN. Firstly, we present the performance of original model based on TN. The phishing identification performance of TSGN, Directed-TSGN, Temporal-TSGN, and Multiple-TSGN, integrating TN, are then displayed respectively. Note that, the concatenated feature vectors are fed into the same classification model for a fair comparison.


\subsubsection{\textbf{TSGN models can enhance phishing account identification ability}}
TABLE~\ref{tab-result} summarises our results in terms of the metric \textit{$F_1$-$Score$}. On the whole, compared with the transaction networks, TSGN, Directed-TSGN, and Temporal-TSGN models indeed have good performance in enhancing the phishing address detection. TSGN proposed as the most basic model is comparable with the five baseline models based on TN, which produces slight improvement under different datasets and feature extraction methods. Line\_Graph2Vec is the improved version of Graph2Vec, which is equivalent to our Graph2Vec w/ TSGN in model design and model performance. 
Directed-TSGN increases the performance of the original phishing identification in 26 out of 30 cases, with a relative improvement of 3.28\% on average. Moreover, since different TSGNs generated by different attribute mapping strategies can capture the different perspectives of a network, one may expect that the fusion of attributes, i.e., Temporal-TSGN, can produce even better detection results. Apparently, the proposed TSGN models improve the identification performance of the TN (original) model in 29 of 30 cases, achieving the optimal result of 98.43\% on EtherG6. Especially, Temporal-TSGN almost always performs best on the six datasets, which incorporating with \emph{Diffpool} outperforms the results based on other baselines.

\subsubsection{\textbf{Transaction feature exploration should be paralleled with topological structure learning}}
It is undisputed that the quality of features extracted from transaction networks will affect the performance of downstream phishing detection model, which is clearly reflected in our experiments. It is obvious that the classification results vary widely based on five different baseline models. \emph{Handcrafted} based on feature assembly is superior to the automatic feature extraction methods relied on learning model, i.e., \emph{Graph2Vec} and \emph{DeepKernel}, with average gains of 9.29\% and 8.26\% respectively. We also find that the handcrafted features elaborately could greatly boost the performance of our TSGN model, achieving a percentage increase of 13.26\% on EtherG1. And then, the experimental results have turned the spotlight on the \emph{Diffpool}. In this case, our TSGN models produce optimal results exceeding those of other methods. Besides, \emph{U2GNN}, which tops the global graph classification leaderboard, did not achieve impressive performances, even underperforming \emph{Handcrafted}. It follows that transaction feature exploiting and topological structure learning of transaction networks are equally important in the scenario of Ethereum phishing identification. The comprehensive and rational exploration of transaction features in the progress of network modeling is the main attribution for the good performance of our TSGNs. 
Furthermore, we find that the complex topological interaction in the transaction networks offers more insights into the model learning of phishing identification. The models on datasets of \emph{Net} type outperform those on datasets of \emph{Star} type in terms of the metric $F_1$-$Score$ in 43 out of 60 cases. While, the models on datasets of \emph{Star} type have great potential for improvement, which produce higher \emph{\%Increase} values than datasets of \emph{Net} type in 11 out of 15 cases. It also shows that the interaction information between 1-hop neighbors could indeed enhance the performance of the detection models.


\begin{table}[h]
\caption{Time consumption (sec.) of constructing TSGN, Directed-TSGN, and Temporal-TSGN.}
\label{tab-time}
\centering
\renewcommand{\arraystretch}{1.5}
\setlength{\tabcolsep}{3mm}{
\begin{tabular}{lrrrr}
    \hline\hline
	Dataset &  TSGN & Directed-TSGN & Temporal-TSGN \\
	\hline
	EtherG1 & 2.42 & 2.35 & 2.29 \\
	EtherG2 & 3.29 & 2.62 & 2.60 \\
	EtherG3 & 605.79 & 93.28 & 71.62 \\
	EtherG4 & 3.58 & 3.45 & 3.31 \\
	EtherG5 & 6.54 & 5.92 & 5.56 \\
	EtherG6 & 801.93 & 202.49 & 165.96 \\
    \hline\hline
\end{tabular}
}
\end{table}

\begin{table*}[h]
\caption{Classification performance of Multiple-TSGN model on multiple-edge transaction networks.}
\label{tab-multiple}
\centering
\renewcommand{\arraystretch}{1.5}
\setlength{\tabcolsep}{4.7mm}{
\begin{tabular}{lcccccc}
    \hline\hline
	Multiple-TSGN &  Handcrafted & Graph2vec & DeepKernel & Diffpool & U2GNN \\
	\hline
	EtherG1 & ${\bf95.00\pm{2.65}}$ & $79.57\pm{3.84}$ & $79.43\pm{3.01}$ & $91.71\pm{2.37}$& $81.14\pm{4.46}$\\
	EtherG2 & ${\bf95.28\pm{2.39}}$ & $76.86\pm{6.99}$ & $79.00\pm{4.04}$ & $91.57\pm{2.07}$& $83.43\pm{3.95}$\\
	EtherG3 & $92.29\pm{2.73}$ & $70.00\pm{4.19}$ & $80.57\pm{6.83}$ & ${\bf95.86\pm{1.49}}$& $81.86\pm{3.20}$\\
	EtherG4 & $92.71\pm{3.16}$ & $74.86\pm{6.13}$ & $78.86\pm{5.34}$ & ${\bf96.71\pm{1.92}}$& $77.86\pm{5.12}$\\
	EtherG5 & ${\bf93.00\pm{2.82}}$ & $73.14\pm{6.18}$ & $78.29\pm{5.74}$ & $90.00\pm{1.92}$& $78.43\pm{5.01}$\\
	EtherG6 & ${\bf91.14\pm{2.70}}$ & $71.43\pm{5.96}$ & $77.00\pm{3.16}$ & $88.57\pm{3.32}$& $79.57\pm{3.32}$\\
    \hdashline
    Avg.Multiple-TSGN    & {\bf93.24}           &74.31           &{\bf78.86}            &92.40         & {\bf80.38} \\
    Avg.TSGNs (best) & 85.88           &{\bf76.62}           &77.26            &{\bf94.90}        & 79.64 \\
    Avg.TN    & 80.36           &70.29           &74.38            &90.47         & 74.69 \\
    \hdashline
    $\%Increase$ (TSGNs) & 8.57\%           &-3.01\%           &2.07\%            &-2.63\%         & 0.93\%  \\
    $\%Increase$ (TN) & 16.03\%           &5.72\%            &6.02\%             &2.13\%         & 7.62\%  \\
    \hline\hline
\end{tabular}}
\end{table*}

\subsubsection{\textbf{TSGNs with attributes can effectively reduce the modeling time of large Ethereum networks}}\label{subsec:complexity}

Furthermore, we record the computational time and compare the time consumption of constructing TSGN, Directed-TSGN, and Temporal-TSGN on six Ethereum datasets. The results are presented in TABLE~\ref{tab-time}, where one can see that the time consumption of constructing TSGN, Directed-TSGN, and Temporal-TSGN only show a slight decrease in these small-scale datasets, such as EtherG1, EtherG2, EtherG4, and EtherG5. However, there is a significant decrease in computational time of Directed-TSGN compared with that of TSGN when considering the larger datasets EtherG3 and EtherG6, decreasing from 6 hundred seconds to less than 93 seconds and from 800 seconds to 200 seconds respectively. In addition, the time consumption of Temporal-TSGN presents a further reduction on the basis of Directed-TSGN. Such results suggest that Directed-TSGN and Temporal-TSGN can further enhance the performance of the algorithm for phishing account identification, while also greatly improving the efficiency of the algorithms.


\begin{figure*}[!t]
	\centering
	\begin{minipage}{0.32\linewidth}
		\centering
		\includegraphics[width=0.9\linewidth]{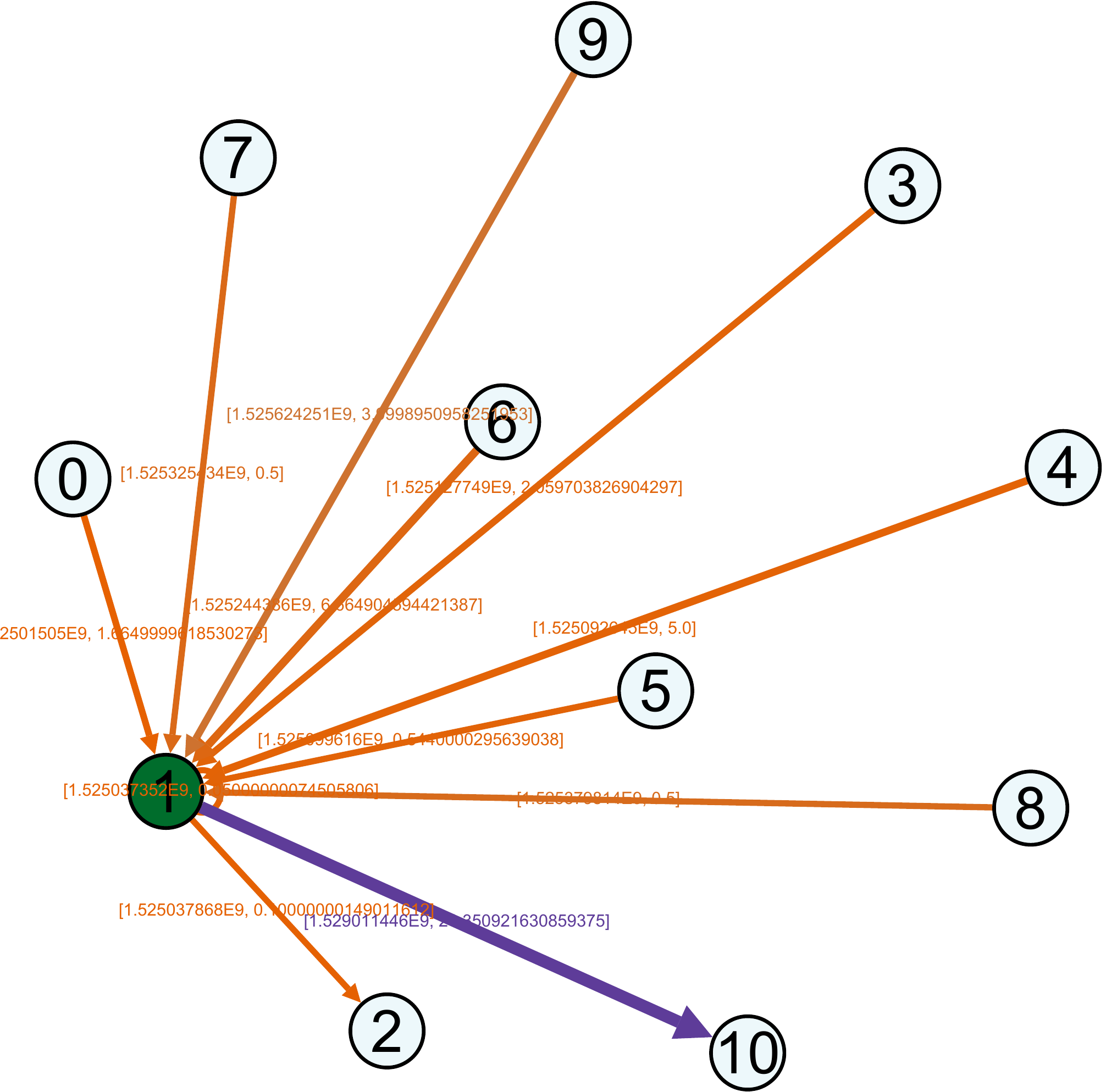}
		\centerline{(a) TN-star}
		\label{subfig:TN-star}
	\end{minipage}
	\centering
	\begin{minipage}{0.32\linewidth}
		\centering
		\includegraphics[width=0.9\linewidth]{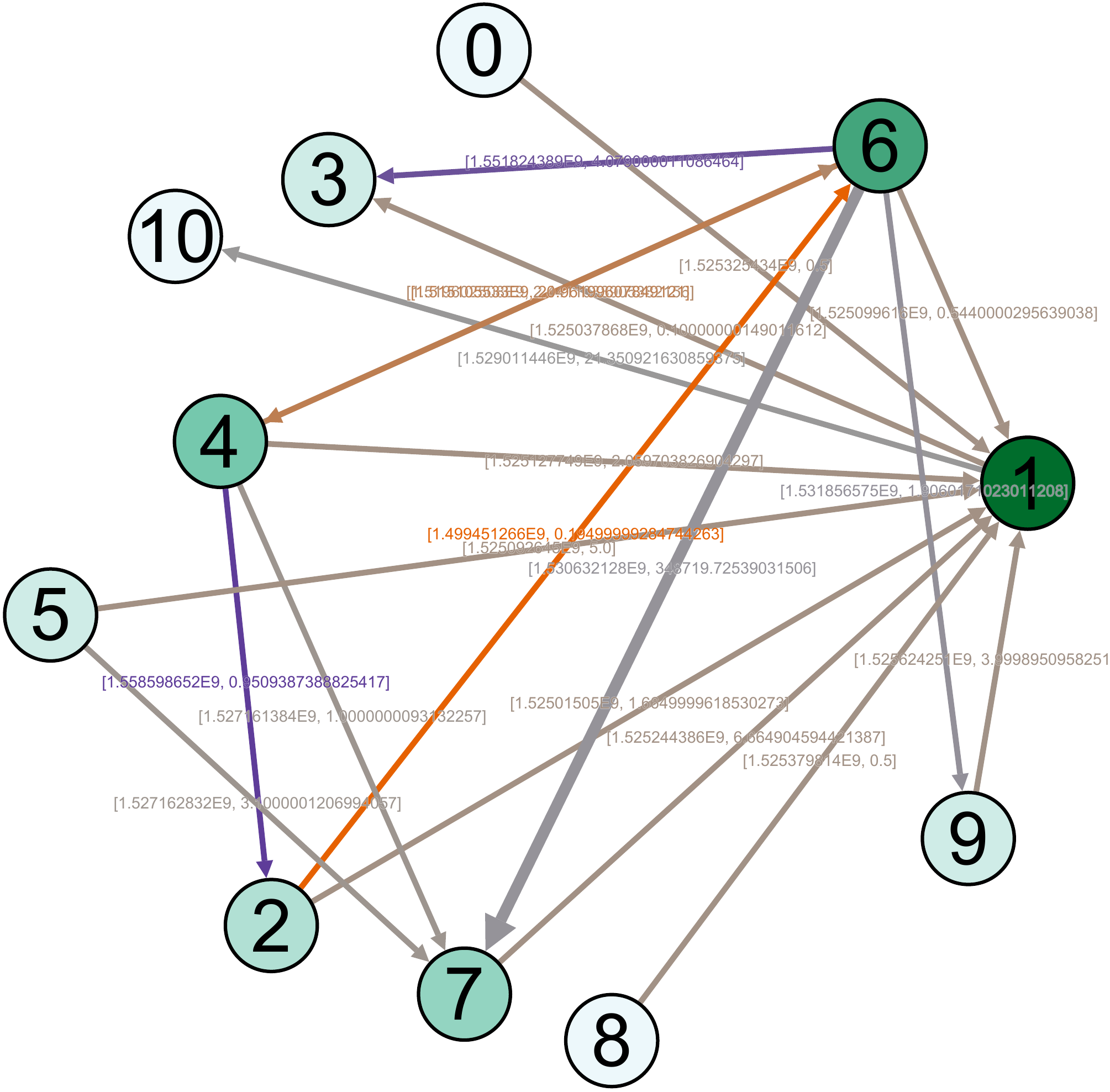}
		\centerline{(b) TN-net}
		\label{subfig:TN-net}
	\end{minipage}
	\centering
	\begin{minipage}{0.32\linewidth}
		\centering
		\includegraphics[width=0.9\linewidth]{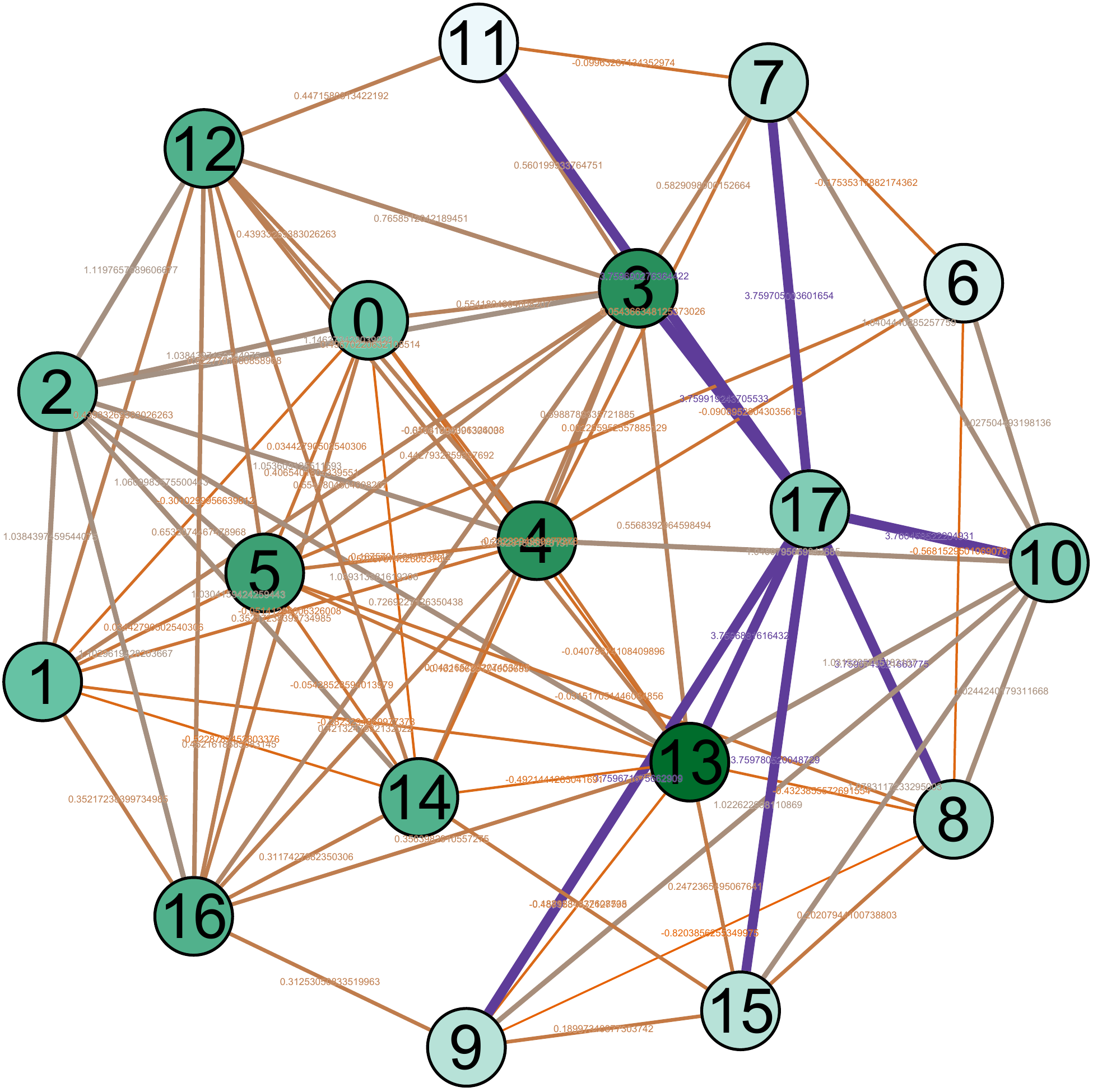}
		\centerline{(c) TSGN}
		\label{subfig:TSGN}
	\end{minipage}

    \centering
	\begin{minipage}{0.32\linewidth}
		\centering
		\includegraphics[width=0.9\linewidth]{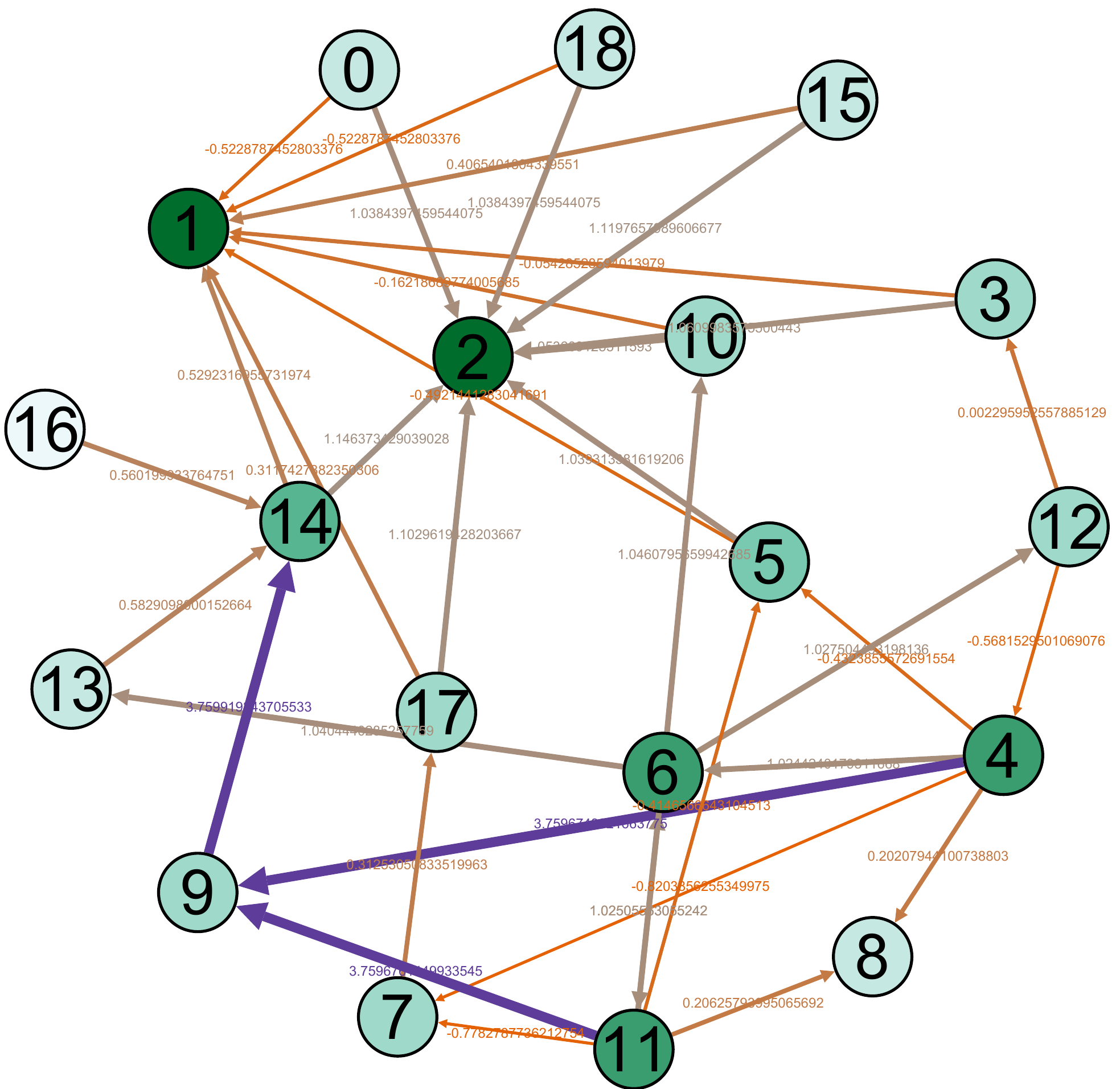}
		\centerline{(d) Directed-TSGN}
		\label{subfig:Directed-TSGN}
	\end{minipage}
	\centering
	\begin{minipage}{0.32\linewidth}
		\centering
		\includegraphics[width=0.9\linewidth]{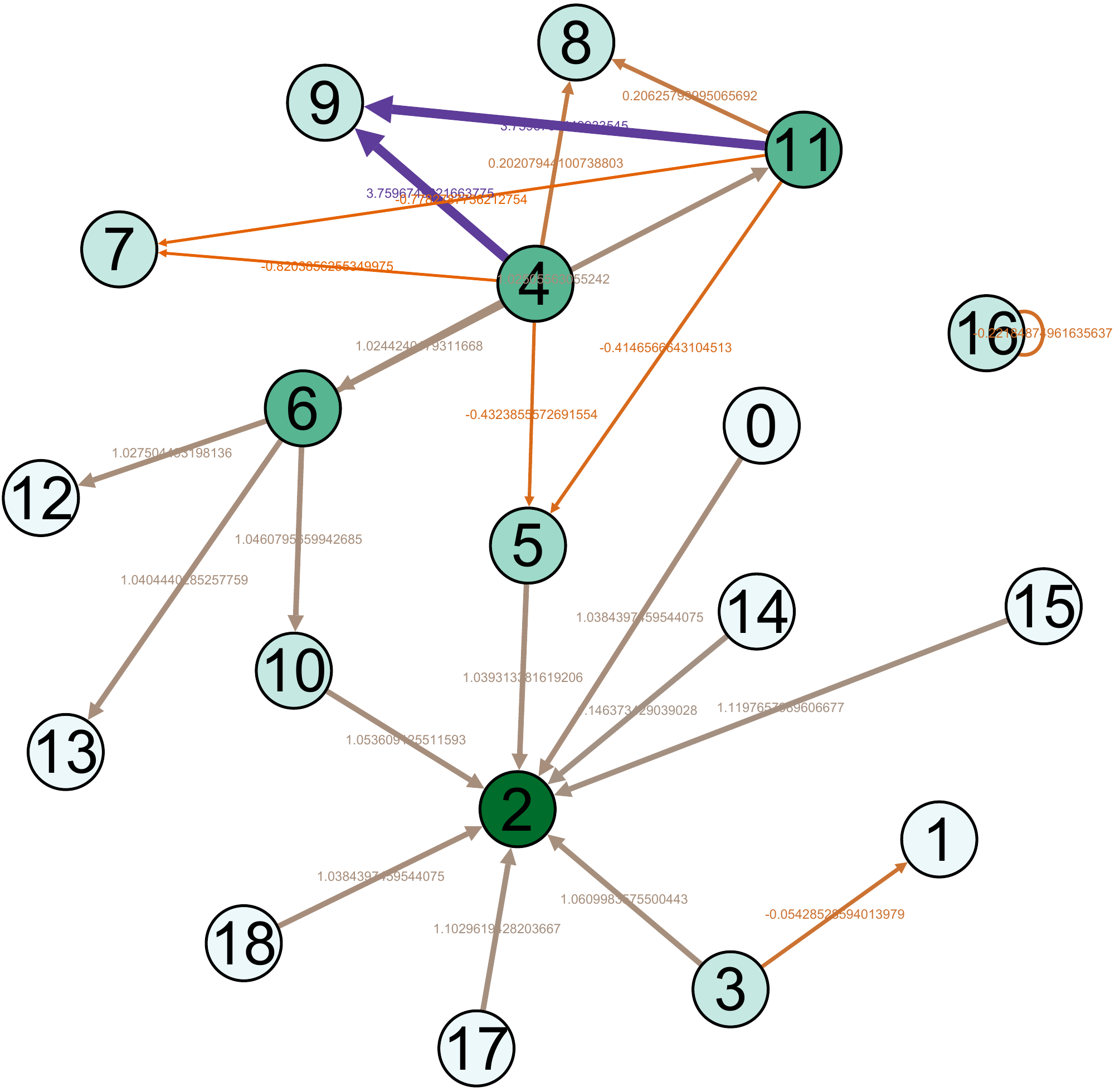}
		\centerline{(e) Temporal-TSGN}
		\label{subfig:Temporal-TSGN}
	\end{minipage}
	\centering
	\begin{minipage}{0.32\linewidth}
		\centering
		\includegraphics[width=0.9\linewidth]{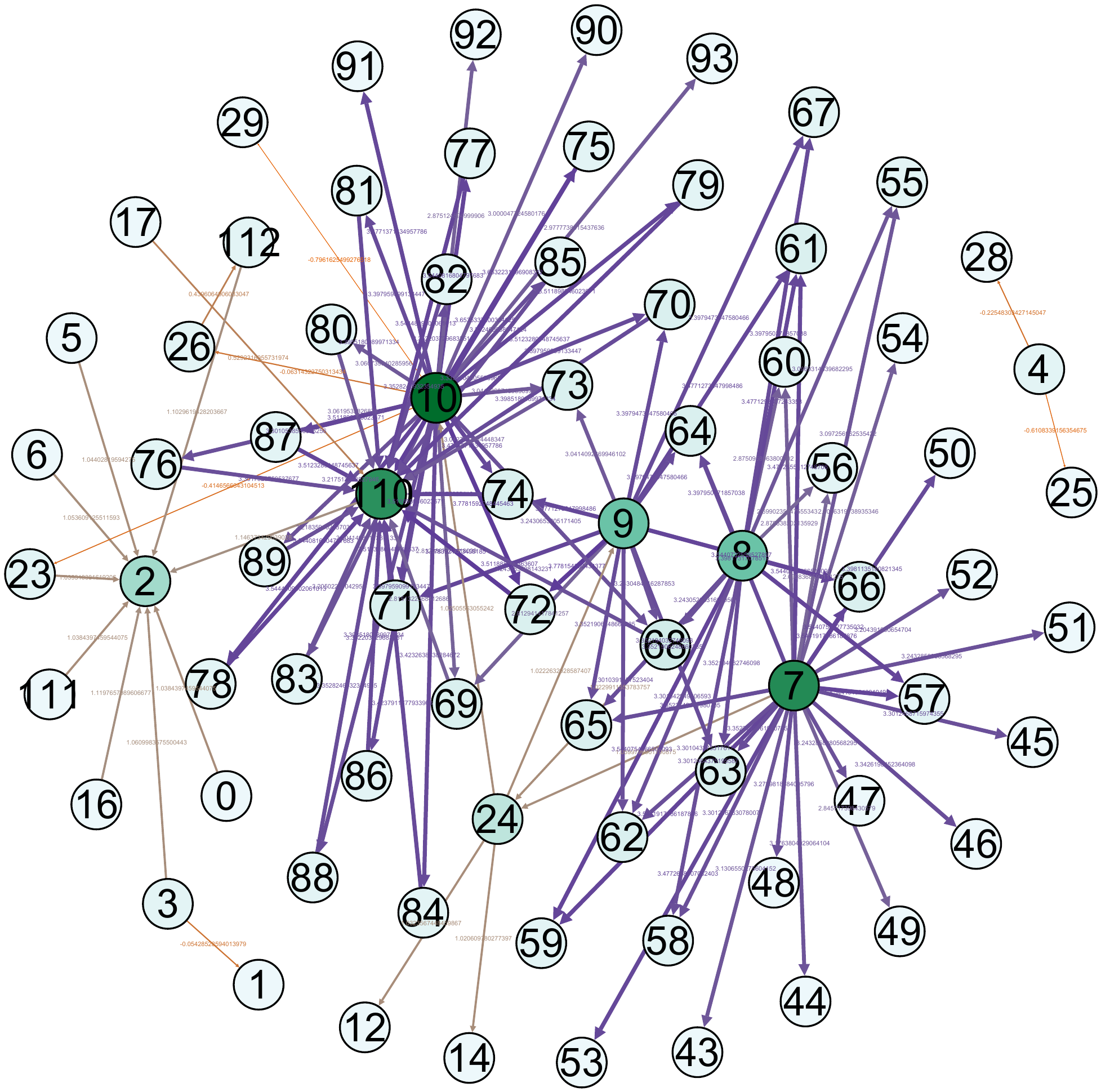}
		\centerline{(f) Multiple-TSGN}
		\label{subfig:Multiple-TSGN}
	\end{minipage}
\caption{Visualization of a real transaction network and its TSGNs.}
\label{fig:datasets-real}
\end{figure*}

\subsubsection{\textbf{Multiple-TSGN performs well on the multi-edge networks}}

Temporal attributes can be adopted to identify and distinguish the multiple same-directed transaction edges between a pair of nodes. Therefore, Multiple-TSGN can dispose of more complex multi-edge transaction networks via extending the scope of application of Temporal-TSGN. In this section, the performance of our TSGN is further evaluated in the case of multi-edge transaction networks, the specific statistics of which are shown in the \emph{Multiedge} collection of TABLE~\ref{tab-data}.
As shown in TABLE~\ref{tab-multiple}, experiments on the six Ethereum datasets show competitive results. We can see that the classification performance of \emph{Graph2Vec}, \emph{DeepKernel}, and \emph{U2GNN} methods on multi-edge datasets does not show much difference compared with the cases as shown in TABLE~\ref{tab-result}. Interestingly, Multiple-TSGN presents its capacity to integrate information of Multi-edge networks under the \emph{Handcrafted} method, achieving good performance of 93.24\% on average, and \emph{Diffpool} still preserves the relatively high detection results of 95.86\% and 96.71\% on EtherG3 and EtherG4, respectively. Besides, compared with TN and TSGNs models, it generates a relative improvement of 16.03\% and 8.57\% on average, respectively, in terms of the $\%Increase$ metric. On the other hand, Multiple-TSGN increases the performance of the original TN model in all cases, with the $\%Increase$ (TN) ranging from 2.13\% to 16.03\%. It follows that our Multiple-TSGN model has a considerable performance when dealing with large-scale transaction networks with complex structural attributes. Overall, TSGN conducted on multi-edge cases can further improve the representation capability of the classification methods.

\subsection{Visualization}\label{subsec:visual}

Moreover, for a better demonstration of network structures, we provide the visualization of real example network and its TSGNs as shown in Fig.~\ref{fig:datasets-real}.
Fig.~\ref{fig:datasets-real} (a)-(b) show the transaction networks of different types, following with TSGN, Directed-TSGN, and Temporal-TSGN shown in (c)-(e) which are mapped from (b) the net-form transaction network. Multiple-TSGN in (f) is constructed based on the multi-edge net-form transaction network.
In these networks, the node with more dark color has a bigger degree value. The arrow on the edge represents the direction of fund flows between accounts, and the thicker edge with higher-level color means a higher transaction amount (orange, gray, and purple are adopted as the color levels in turns). It can be found that Directed-TSGN can indeed reduce the scale of TSGN and yield a clear structure of transfer flows, and Temporal-TSGN further filters the network structure of Directed-TSGN by the transaction temporal attribute, where some edges disappear such as (9,14), (16,14) and (14,1) and forth. Multiple-TSGN is much more complex than the other TSGNs, providing more interaction information from different accounts around the target node, benefiting the phishing detection.

\begin{figure}[ht]
	\centering
	\includegraphics[width=1\linewidth]{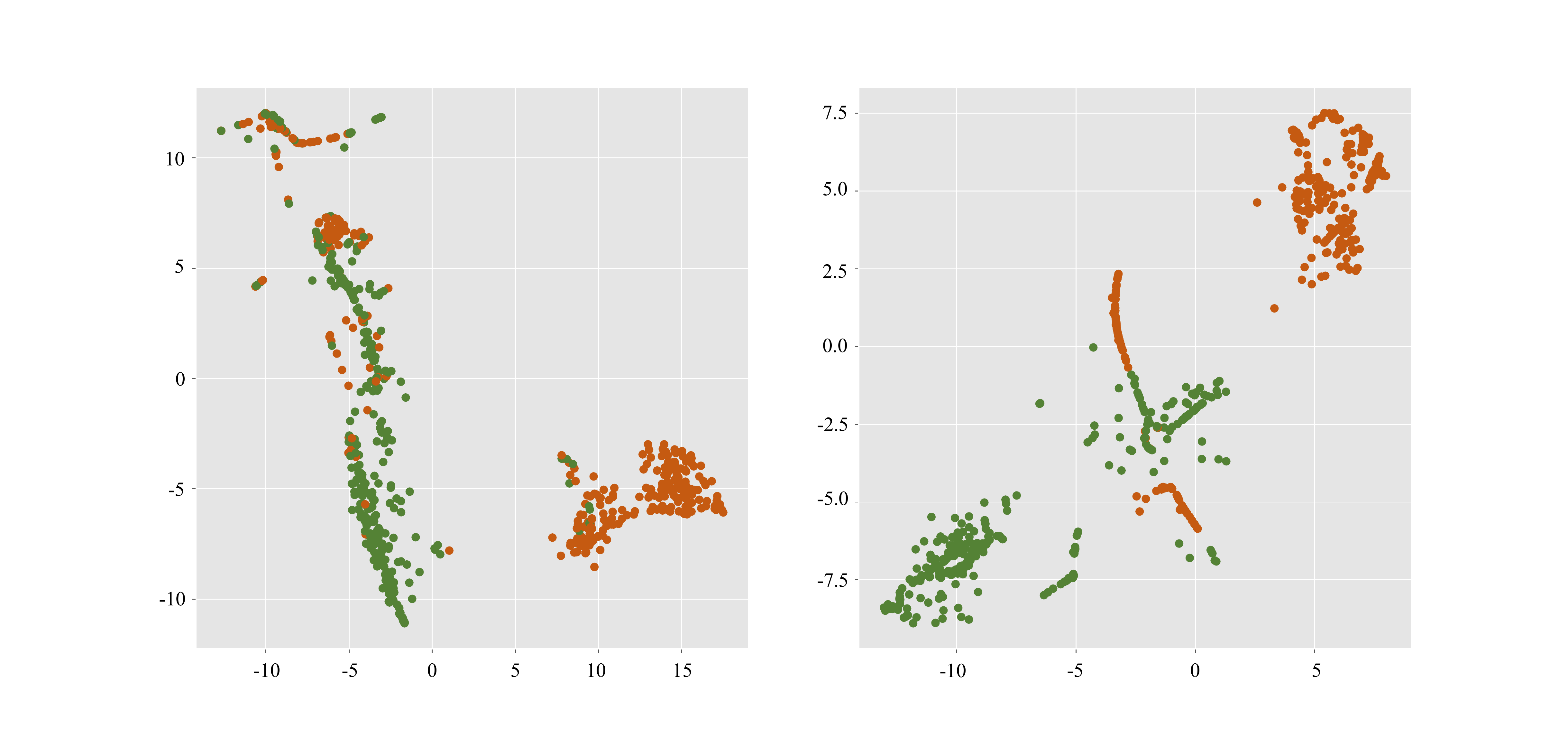}
	\caption{The t-SNE visualization of structural features using Diffpool with (left) TN and with (right) Directed-TSGN. The same color of points represent the same class of graphs in EtherG1 dataset.}
	\label{fig:visualization}
\end{figure}

As a simple case study, we utilize the t-SNE to visualize the results of classification on EtherG1 dataset based on Diffpool method to verify the effectiveness of our TSGN model. Here, we choose the Directed-TSGN to visualize since it produces the good performance that enhances the detection performance of Diffpool most, with $\%Increase$ of 14.58\%.
As shown in Fig.~\ref{fig:visualization}, the structural features are located in different places by utilizing t-SNE. The left shows the original detection result using Diffpool with TN model, while the right depicts the optimized distribution of the same dataset using Diffpool with Directed-TSGN. One can see that the transaction networks in EtherG1 dataset can indeed be distinguished by the original features of Diffpool, but it appears that the distinction of networks could become more explicit after TSGN mapping with direction, demonstrating the effectiveness of our Directed-TSGN model. It also suggests that our TSGN model provides more predominant features contrasted with original TN model.

\section{Conclusion} \label{sec:conclusion}


In this paper, we propose a novel transaction subgraph network (TSGN) model for phishing detection. By introducing different mapping mechanisms into the transaction networks, we built TSGN, Directed-TSGN, Temporal-TSGN models to enhance the classification algorithms. 
Compared with the baselines only based on the TN, our TSGNs indeed provide more potential information to benefit the phishing detection. 
By comparing with the TSGN, Directed-TSGN and Temporal-TSGN are of the controllable scale and indeed have much lower time cost, benefiting the network feature extraction methods to learn the network structure with higher efficiency.
Experimental results manifest that, combined with network representation algorithms, these TSGN models can capture more features to enhance the classification algorithm and improve phishing nodes identification accuracy in Ethereum. 
In particular, when integrating the appropriate feature representation methods, such as Handcrafted and Diffpool, TSGNs can achieve the best results on all datasets.

In the future, we will extend and improve our model for suit more blockchain data mining tasks, such as Pondzi scheme detection, transaction tracking, etc.

\ifCLASSOPTIONcompsoc
  \section*{Acknowledgments}
\else
  \section*{Acknowledgment}
\fi
This work was partially supported by the Key R\&D Program of Zhejiang under Grant No. 2022C01018, by the National Natural Science Foundation of China under Grant No. U21B2001 and 61973273, by the Zhejiang Provincial Natural Science Foundation of China under Grant No. LR19F030001, and by the National Key R\&D Program of China under Grant No. 2020YFB1006104.

\ifCLASSOPTIONcaptionsoff
  \newpage
\fi

\bibliographystyle{IEEEtran}
\bibliography{bare_adv}

\end{document}